\documentclass[aps,prb,preprint,superscriptaddress]{revtex4-2}

\usepackage{dcolumn}   
\usepackage{bm}        
\usepackage{amsmath}
\usepackage{amssymb}
\usepackage[utf8]{inputenc}
\usepackage{graphicx,siunitx}
\usepackage{amsmath}
\DeclareUnicodeCharacter{2212}{-}
\usepackage{booktabs}
\usepackage[svgnames]{xcolor}
\usepackage[export]{adjustbox}
\usepackage[percent]{overpic}
\usepackage{array}
\usepackage[style=base]{caption}
\DeclareSIUnit\angstrom{\text {Å}}
\newcommand*{\vcenteredhbox}[1]{\begingroup
	\setbox0=\hbox{#1}\parbox{\wd0}{\box0}\endgroup}

\begin{document}
	\title{Soft mode induced structural phase transition in Ba$_2$ZnTeO$_6$ at high pressure}
	
    \author{Bidisha Mukherjee}
	\affiliation{Department of Physical Sciences, Indian Institute of Science Education and Research Kolkata, Mohanpur Campus, Mohanpur 741246, Nadia, West Bengal, India.}
	\affiliation{National Centre for High-Pressure Studies, Indian Institute of Science Education and Research Kolkata, Mohanpur Campus, Mohanpur 741246, Nadia, West Bengal, India.}
    \author{Surajit Adhikari}
        \affiliation{Department of Physics, School of Natural Sciences, Shiv Nadar Institution of Eminence, Greater Noida, Gautam Buddha Nagar, Uttar Pradesh 201314, India.}
    \author{Mrinmay Sahu}
	\affiliation{Department of Physical Sciences, Indian Institute of Science Education and Research Kolkata, Mohanpur Campus, Mohanpur 741246, Nadia, West Bengal, India.}
	\affiliation{National Centre for High-Pressure Studies, Indian Institute of Science Education and Research Kolkata, Mohanpur Campus, Mohanpur 741246, Nadia, West Bengal, India.}
    \author{Asish Kumar Mishra}
	\affiliation{Department of Physical Sciences, Indian Institute of Science Education and Research Kolkata, Mohanpur Campus, Mohanpur 741246, Nadia, West Bengal, India.}
	\affiliation{National Centre for High-Pressure Studies, Indian Institute of Science Education and Research Kolkata, Mohanpur Campus, Mohanpur 741246, Nadia, West Bengal, India.}
	\author{Bhagyashri Giri}
	\affiliation{Department of Physical Sciences, Indian Institute of Science Education and Research Kolkata, Mohanpur Campus, Mohanpur 741246, Nadia, West Bengal, India.}
	\affiliation{National Centre for High-Pressure Studies, Indian Institute of Science Education and Research Kolkata, Mohanpur Campus, Mohanpur 741246, Nadia, West Bengal, India.}
    \author{Priya Johari}
        \affiliation{Department of Physics, School of Natural Sciences, Shiv Nadar Institution of Eminence, Greater Noida, Gautam Buddha Nagar, Uttar Pradesh 201314, India.}
    \author{Konstantin Glazyrin}
        \affiliation{Photon Science, Deutsches Elektronen Synchrotron, 22607 Hamburg, Germany}
    \author{Goutam Dev Mukherjee}
	\email [Corresponding author: ]{goutamdev@iiserkol.ac.in}
	\affiliation{Department of Physical Sciences, Indian Institute of Science Education and Research Kolkata, Mohanpur Campus, Mohanpur 741246, Nadia, West Bengal, India.}
	\affiliation{National Centre for High-Pressure Studies, Indian Institute of Science Education and Research Kolkata, Mohanpur Campus, Mohanpur 741246, Nadia, West Bengal, India.}
	\date{\today}
	
	\begin{abstract}
  In this paper, we present a thorough investigation of vibrational, structural,  and electronic properties of perovskite-type rhombohedral Ba$_2$ZnTeO$_6$ (BZTO) under systematic application of pressure. To carry out the analysis, we have performed pressure-dependent Raman spectroscopic measurements, synchrotron XRD, and density functional theory-based calculations. At ambient conditions, BZTO stabilizes in $R\bar{3}m$ space group, which under pressure undergoes a structural transition to a monoclinic phase with space group $C2/m$ at around 18~GPa. In-depth Raman analysis reveals softening of a phonon mode E$_g$ ($\sim $ 28cm$^{-1}$) leads to the structural phase transition. First principle DFT calculations also indicate that the doubly degenerate soft mode associated with the in-phase TeO$_6$ octahedral rotation drives the structure to a lower symmetry phase $C2/m$.   
   
\end{abstract} 
\maketitle
	
\section{INTRODUCTION}
	
Perovskite oxides are compounds with the chemical formula ABO$_3$, where A is the larger cation and B is the smaller cation, which ideally have cubic structure with $Pm3m$ space group. Compounds of this class are of huge technological interest due to their excellent potential in catalysis \cite{Pena2001,D0TA09756J,GRABOWSKA201697, kumar2020,Zhu2014} and ferroelectric applications \cite{Cohen1992}. These compounds present exceptional tunability due to the large number of A and B cation combinations. This variety further increases manyfold in double perovskite oxides (DPOs) with the general form A$_2$BB$^{\prime}$O$_6$ (where, A represents alkaline earth or rare earth materials having larger cation, B and B$^{\prime}$ represent smaller cation). They possess a variety of desirable physical characteristics due to the different possible combinations of B and B$^{\prime}$ cations. Hence, external perturbation-induced structural transitions in such materials are of technological interest. DPOs are reported to possess multiple magnetic and electrical properties such as high thermoelectric figure-of-merit \cite{tanwar2019enhancement}, catalytic properties \cite{xu2019double}, etc. Among the limited number of studies that have investigated on the effects of pressure at room temperature on DPOs, Ba$_2$YTaO$_6$ is reported to undergo structural transition from $Fm\bar{3}m$ to $I4/m$ at 5.6~GPa \cite{Ba2YTaO6}. Ba$_2$BiSbO$_6$ undergoes a pressure-induced structural transition from rhombohedral ($R\bar{3}$) to monoclinic ($I2/m$) above 4~GPa \cite{Ba2BiSbO6}. Rhombohedral ($R\bar{3}$) to monoclinic ($I2/m$) transition has also been found in Ba$_2$BiTaO$_6$ either by lowering the temperature to 17~K or by increasing the pressure up to 4~GPa.

The compounds Ba$_2$M$^{2+}$Te$^{6+}$O$_6$ (BMTO), where M$^{2+}$ = Co, Ni, Cu, or Zn feature a structure of face-sharing TeO$_6$ and MO$_6$ octahedra, connected by corner-sharing TeO$_6$ octahedra. Ba$_2$CoTeO$_6$ has octahedral doublets \cite{Ba2CoTeO6}, whereas the rest contain octahedral triplets. Ba$_2$CuTeO$_6$ (BCuTO) adopts a distorted 12L (12-layered)-monoclinic structure ($C2/m$) \cite{BCuTOSurajitSaha}, while Ba$_2$NiTeO$_6$ (BNTO) and Ba$_2$ZnTeO$_6$ (BZTO) both form a 12L-rhombohedral structure ($R\bar{3}m$ space group) at room temperature \cite{BNTO_magneticordering,BZTOSurajitSaha}. This difference in crystal structure of BCuTO from BZTO and BNTO are coming from the presence of Jahn-teller active Cu$^{2+}$ ion in BCuTO. The low temperature studies on these materials are thoroughly explored and show promising results. Additionally, BZTO and BNTO have comparable lattice parameters, making them sister materials. Earlier studies have shown that lowering the temperature down to even 1.8~K, does not produce a structural transition in BNTO \cite{BNTO_SurajitSaha}. But, BZTO surprisingly  transitions to a lower symmetry phase ($C2/m$) when the temperature is lowered to 150~K \cite{BZTOSurajitSaha}. Unlike BCuTO, BZTO has non-magnetic Zn$^{2+}$ ion which is Jahn-Teller inactive and typically causes its complexes to retain their symmetry. Studies have shown that lowering in symmetry with lowering temperature in BZTO is caused by a softening of E$_g$ mode accompanied by an in-plane elastic distortion. All of these results have established the importance of investigating and understanding the structural complexity of this type of less common hexagonal perovskites. However, the behaviour of BMTOs in low temperature regime has been thoroughly studied. But there are no such studies on high pressure regions on BMTOs which we believe will provide better understanding and more clarity on the lattice dynamics of BMTOs. Application of controlled pressure can modify atomic arrangements and, consequently, the physical properties of these materials, which might be of technological relevance. Applying pressure on BaTeO$_3$, one of the end-single-perovskite members of BZTO, leads to abrupt changes in electrical transport measurements and irreversible structural transition at 12.7~GPa \cite{BaTeO3}. As mentioned earlier, BZTO stabilizes in $R\bar{3}m$ space group and low temperature induces lower symmetry in BZTO. On the other hand, the high-pressure-high-temperature synthesis process stabilizes it to the cubic phase \cite{BZTOmoreiraPRM}. All these facts make the structural dynamics of BZTO very rich and interesting to study further. Thus the unique and intriguing properties of Ba$_2$M$^{2+}$Te$^{6+}$O$_6$ family under different temperature conditions, which are yet unexplored in high-pressure conditions, make us curious to investigate the structural and electronic properties of the BZTO sample under systematic application of pressure. Understanding the detailed mechanisms behind interactions among structural, electronic properties is crucial for tailoring materials for specific applications in optoelectronics, photovoltaics, and beyond.

In this study, we investigate the vibrational, structural, and electronic properties of BZTO. A pressure-dependent Raman spectroscopy up to 40.3~GPa reveals sudden slope changes of the peak centers and FWHM of the Raman modes in between the pressure range of 7.5~GPa and 10~GPa. The increase in trigonality of the structure as well as the sudden increase in bond angle variance of the TeO$_6$ octahedra confirm that the anomalies observed in the Raman modes are due to the structural instability developed in the parent structure upon compression. We performed an in-depth analysis of the behavior of low-frequency phonon mode with increasing pressure which indicates a lattice instability of the ambient phase that starts around 10~GPa and consequently leads to a structural transition around 18~GPa. A pressure-dependent XRD measurement shows the appearance of new peaks around 20~GPa confirming a structural transition from rhombohedral to monoclinic phase. Further studies involving DFT calculations show that the unstable low-frequency phonon mode drives the rhombohedral to monoclinic phase transition.

	\section{EXPERIMENTAL SECTION}
We have used high-purity raw materials to synthesize BZTO powder using the solid-state synthesis method as described by Badola et al. \cite{BZTOSurajitSaha}.
Its phase purity was verified using Rigaku X-ray diffractometer having Cu-K$\alpha$ source with wavelength of 1.5406~\AA. We performed Energy Dispersive X-ray spectroscopy (EDX) measurements to examine the stoichiometric ratio of the synthesized powder. The grain size of the sample was recorded using SUPRA 55 VP-
4132 CARL ZEISS Field Emission Scanning Electron Microscopy (FESEM) running at 5.21 KeV. ImageJ software \cite{Schneider2012} was used to analyze the data.

High-pressure Raman spectroscopic measurements were performed using a piston-cylinder type diamond anvil cell (DAC) having a culet of \SI{300}{\micro\metre} in diameter. A \SI{290}{\micro\metre} thick steel gasket was indented to a thickness of \SI{45}{\micro\metre}. At the center of the indented region, a hole of diameter \SI{100}{\micro\metre} was drilled through the gasket using an electric discharge machine. The gasket was then placed on the lower diamond. A minimal amount of sample, pressure transmitting medium (PTM), and Ruby chips of approximately \SI{5}{\micro\metre} size were loaded into the central hole. 4:1 methanol-ethanol mixture was used as the PTM and the Ruby fluorescence technique was used for pressure calibration \cite{mao1986calibration}.

The Raman spectra of the sample were taken with a Monovista confocal micro-Raman system from S\&I GmbH equipped with a Cobolt Samba \SI{532}{\nano\metre} diode-pumped laser. A long working distance infinitely corrected $20\times$ objective was used to focus the laser beam and to collect the scattered signal from the sample using back-scattering geometry. The laser spot size on the sample surface was around \SI{4}{\micro\metre}. The collected light was dispersed using a grating with 1500 grooves/mm having a spectral resolution of about~\SI{1.2}{\centi\metre}$^{-1}$. An edge filter for Rayleigh line rejection was used which has a cut-off near ~\SI{80}{\centi\metre}$^{-1}$. For a detailed analysis of the low-frequency Raman mode, a Bragg filter with cut-off ~\SI{4}{\centi\metre}$^{-1}$ was used.

High-pressure XRD data was taken at PETRA III, P02.2 beamline, Germany using a monochromatic X-ray beam of wavelength 0.2907~Å and spot size 8$\times$3\SI{}{\micro\metre}$^2$. Perkin Elmer (XRD1621) detector was used to collect the 2D diffraction images. For high pressure XRD measurements we used Rhenium gasket and Neon gas as PTM. Then we calibrated the distance from the sample to the detector using the XRD pattern of a standard sample CeO$_2$. Diffraction patterns were integrated to $2\theta$ versus intensity profile using DIOPTAS software \cite{dioptas}. To analyse the XRD data, GSAS \cite{toby2001expgui}, EoSfit7 \cite{gonzalez2016eosfit7}, and Vesta \cite{momma2008vesta} software were used.
Electronic absorption spectra were recorded on a Shimadzu UV-Vis-NIR spectrometer.
\section{Computational Details}

To support experimental results, we also performed first-principles density functional theory (DFT) \citep{chapter2-36,chapter2-37} calculations using projector-augmented wave (PAW) pseudopotentials \citep{chapter1-33}, as implemented in Vienna Ab initio Simulation Package (VASP) \citep{chapter1-31,chapter1-32}. The PAW pseudopotentials with valence-electron configurations considered for Ba, Zn, Te, and O were 5s$^{2}$5p$^{6}$6s$^{2}$, 4s$^{2}$3d$^{10}$, 5s$^{2}$5p$^{4}$, and 2s$^{2}$2p$^{4}$, respectively. For structural optimization, the Perdew-Burke-Ernzerhof (PBE) exchange-correlation (xc) functional, based on the generalized gradient approximation (GGA) \citep{chapter1-34}, was used to model electron-electron interactions. The plane-wave cutoff energy was set to 520 eV, with the convergence criterion for the electronic self-consistent field iterations established at 10$^{-6}$ eV. The lattice constants and atomic positions were fully optimized until the Hellmann-Feynman forces \cite{hellmann2015hans, feynmanforces} on each atom were minimized to less than 0.01 eV/Å. For both the phases we have used experimental lattice parameters as the starting model for doing the DFT calculations. The crystal structures were optimized utilizing a $\Gamma$-centered $4\times4\times4$ $\mathbf{k}$-point sampling scheme for Brillouin zone integration. Subsequently, the optimized structures were visualized using the Visualization for Electronic and STructural Analysis (VESTA) software package \citep{momma2008vesta}. The phonon spectra were calculated using the density functional perturbation theory (DFPT) method \citep{chapter1-60} with $2\times2\times2$ supercells, as implemented in the PHONOPY package \citep{chapter3-6}. In case of calculations of electronic properties, the lattice parameters are fixed to their experimental values. Only the internal degrees of freedom are allowed to relax. The electronic properties of all the structures were investigated by calculating the atom-projected partial density of states (pDOS) and the electronic band structure using the PBE xc functional. The band structure at ambient conditions was also carried out using hybrid HSE06 xc functional for more accurate estimation \cite{hybrid_HSE06}.

\section{RESULTS AND DISCUSSION}

	\subsection{Sample Characterization}
\begin{figure}[h]	
    \hspace{-5cm}(a) \hspace{0.5\linewidth} (b) \\
    \vcenteredhbox{\includegraphics[width=0.5\linewidth]{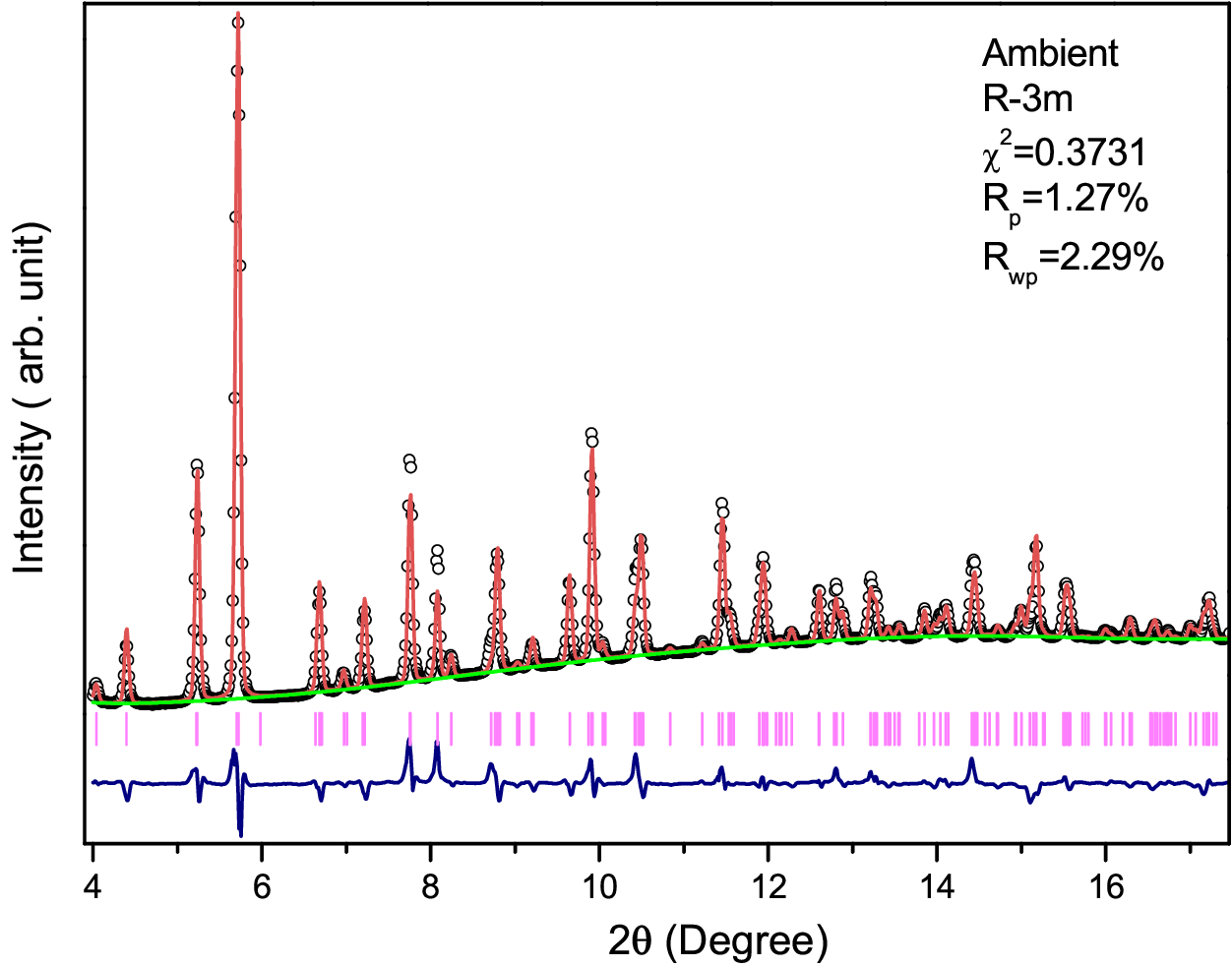}} \hspace{1cm} \vcenteredhbox{\includegraphics[width=0.27\linewidth]{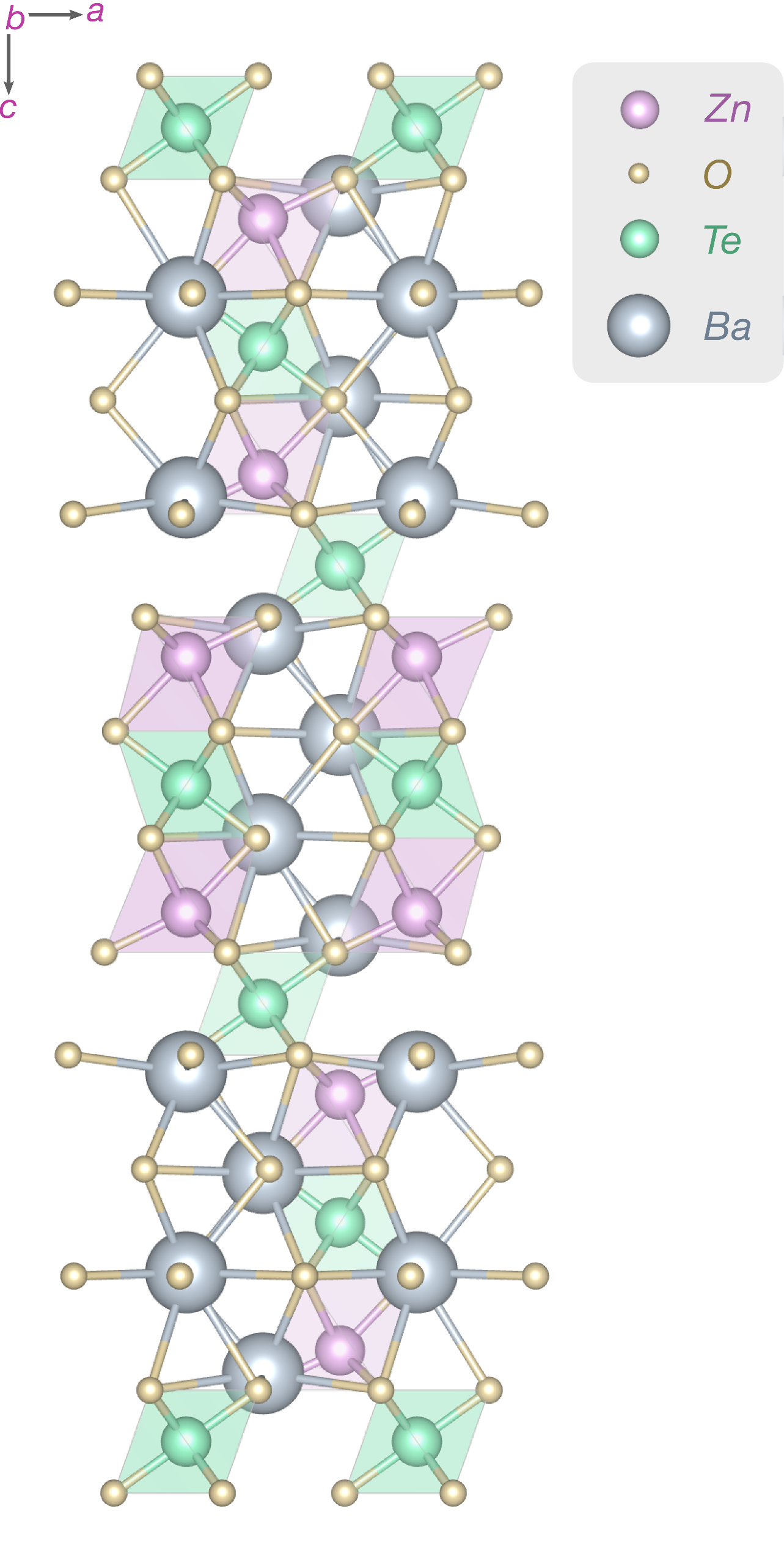}}
    \\ \hspace{-5cm}(c) \hspace{0.5\linewidth} (d) \\
     \vcenteredhbox{\includegraphics[width=0.49\linewidth]{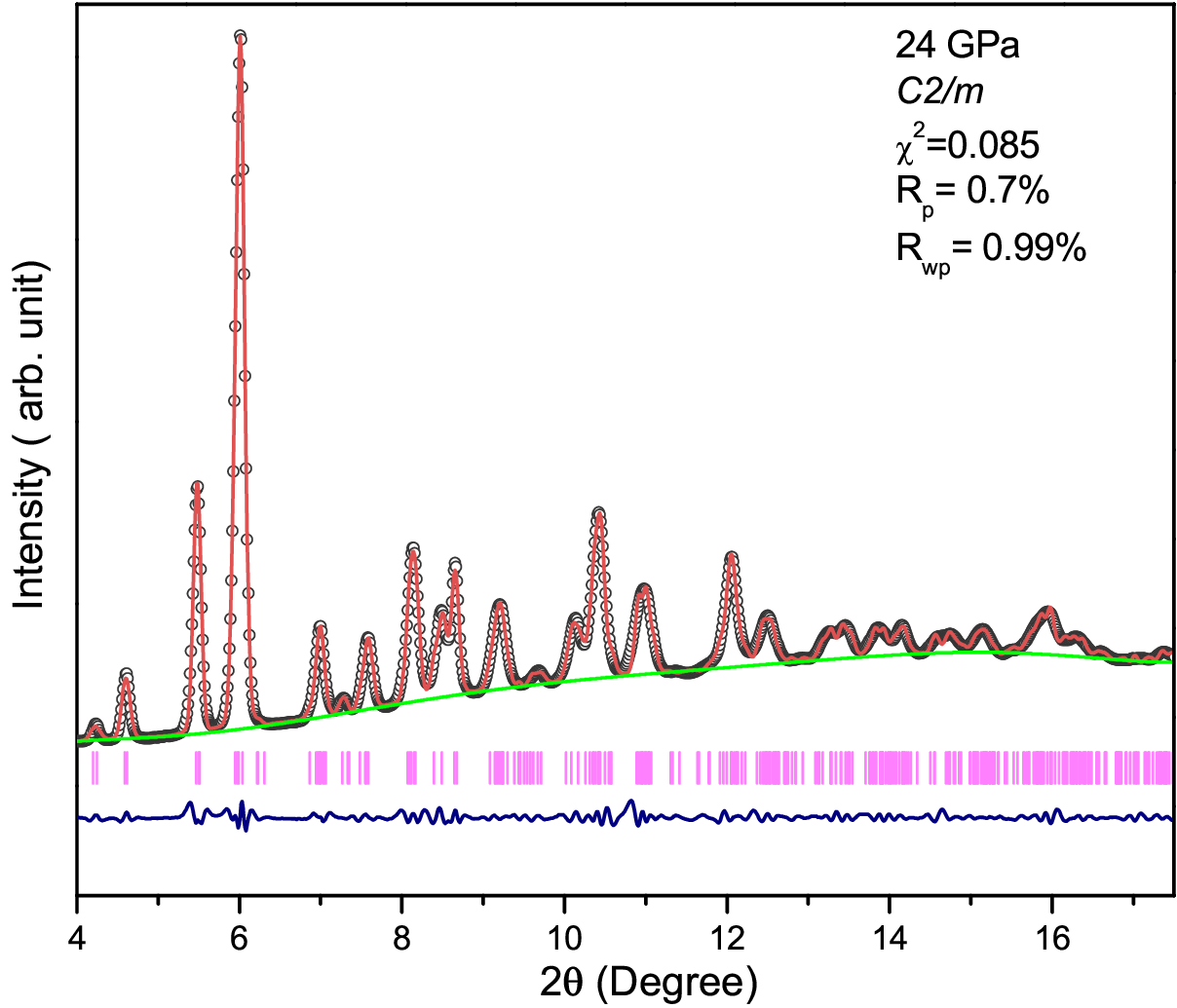}} \hspace{1cm}\vcenteredhbox{\includegraphics[width=0.28\linewidth]{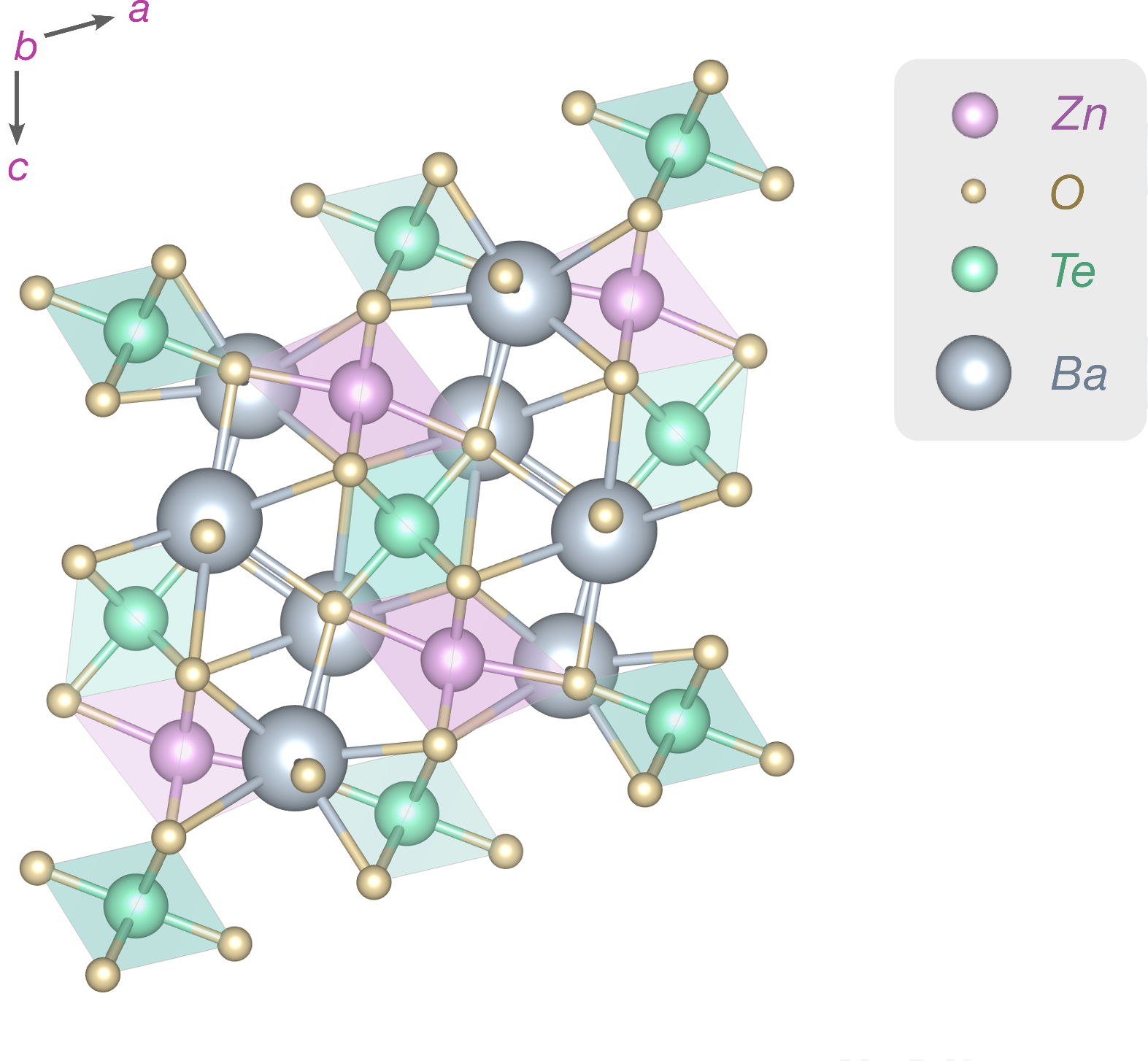}}
		\caption{(a), (b) Rietveld refinement of the XRD pattern at ambient conditions and 24.1~GPa respectively. Black hollow circles represent experimental data. Red, green, and navy lines are Rietveld fit to the experimental data, background, and difference between experimental and calculated data, respectively. The magenta vertical lines show the Bragg peaks of the sample. Schematic representation of the unit cell at (c) ambient conditions and (d) at 24.1~GPa are shown in the right side. }
	\label{fig:XRD}
\end{figure}

The powder XRD pattern of the synthesized BZTO sample is shown in FIG.~\ref{fig:XRD}. The Bragg lines of the ambient XRD data can be indexed using $R\bar{3}m$ space group and the refined lattice constants are $a = 5.8235(1)~\si{\angstrom}$, $c =  28.6919(1)~\si{\angstrom}$ and volume, $V = 842.7029(1)~\si{\angstrom}^3$ which are in a good agreement with the literature \cite{BZTOSurajitSaha,BZTOmoreiraPRM}. For Rietveld refinement, we used the atomic positions of BNTO \cite{BNTO_SurajitSaha}, a sister material of BZTO, as initial atomic positions. The Rietveld refinement at ambient conditions is shown in FIG.~\ref{fig:XRD}(a) which shows an excellent fit with R$_p=1.27\%$ and R$_{wp}=2.29\%$ and the refined relative atomic positions are tabulated in Table~\ref{table:2}.
\begin{table}[ht!] 
	\begin{center}
		\begin{tabular}{ |c|c|c|c|c|c| } 
			\hline
			Atom & Wyckoff position & x/a & y/b & z/c\\
			\hline
			Ba(1) & 6c & 0.0 & 0.0 & 0.1288(2)\\
			Ba(2) & 6c & 0.0 & 0.25 & 0.2815(1)\\
			Zn & 6c & 0.0 & 0.0 & 0.4025(3)\\
			Te(1) & 3a & 0.0 & 0.0 & 0.00\\
			Te(2) & 3b & 0.0 & 0.0 & 0.5\\
			O(1) & 18h & 0.153 & -0.153 & 0.4593\\
			O(2) & 18h & 0.177 & -0.177 & 0.6275\\
			\hline
		\end{tabular}\\
	\end{center}
	\caption{Relative atomic positions at ambient condition after Rietveld refinement}
	\label{table:2}
\end{table} 
The unit cell of BZTO is 12L (12 layers) type which is formed with face sharing and corner sharing ZnO$_{6}$ and TeO$_6$ octahedra as shown in FIG.~\ref{fig:XRD}(b). We observed the grain size of the microstructure of BZTO powder using FESEM and estimated the grain size from the different positions of the sample. The average grain size measured using ImageJ software is~\SI{0.7}{\micro\metre}. Two images taken from different positions of the sample are shown in FIG.~S1.~(a) and (b)~\cite{supp}. The elemental analysis was done by EDX measurement which results in a good agreement with the stoichiometric ratio and is shown in Supplementary FIG.~S1.~(c)~\cite{supp}.

\subsection{High Pressure Raman Spectroscopic Study}
\begin{figure}[h]
	\centering
	\includegraphics[width=0.8\linewidth]{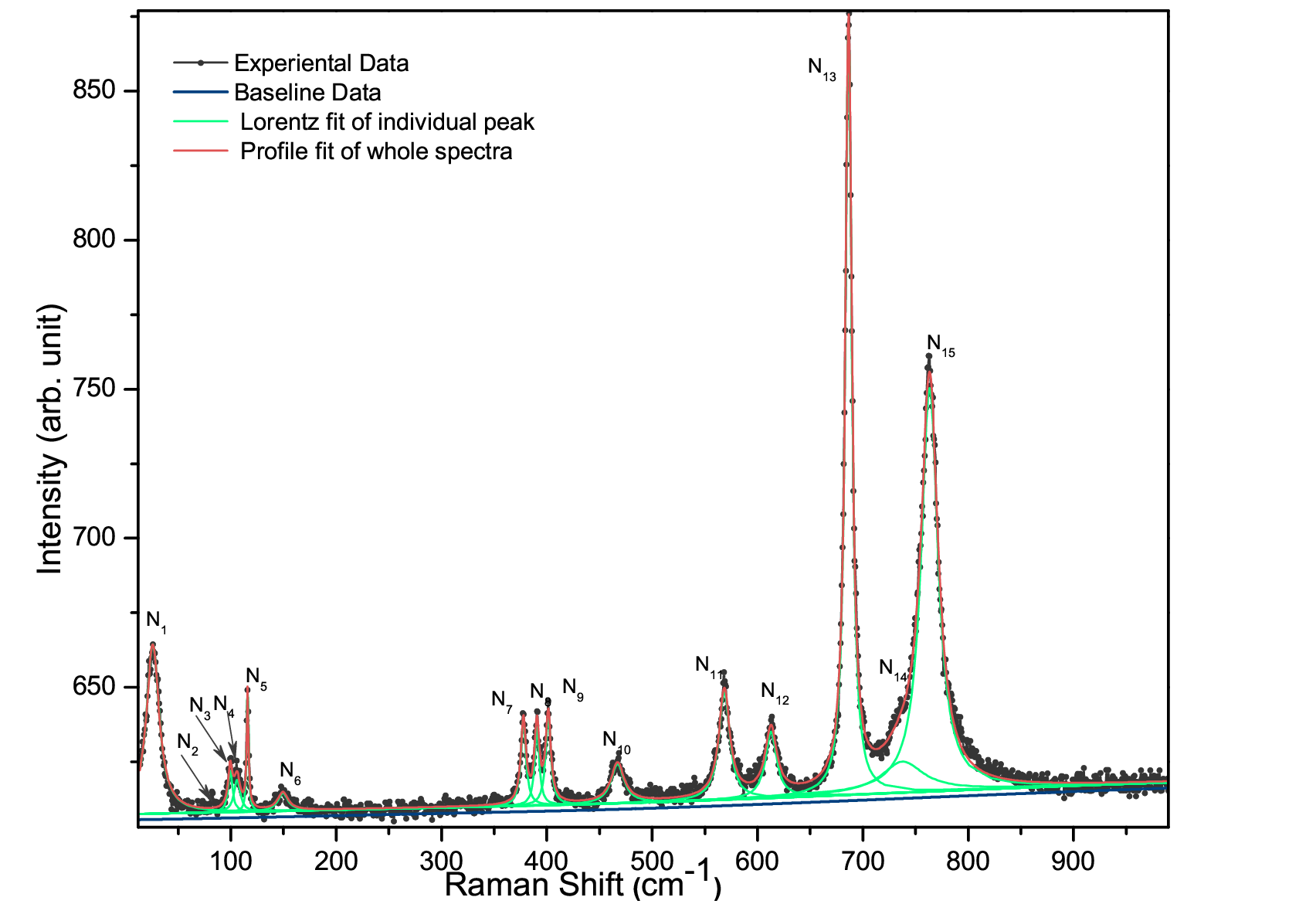}
	\caption{Raman spectra of BZTO collected at ambient conditions. The background corrected spectrum is fitted using the Lorentzian profile.}
	\label{fig:AmbRaman}
\end{figure}

\begin{figure}
	\centering
	\includegraphics[width=0.24\linewidth]{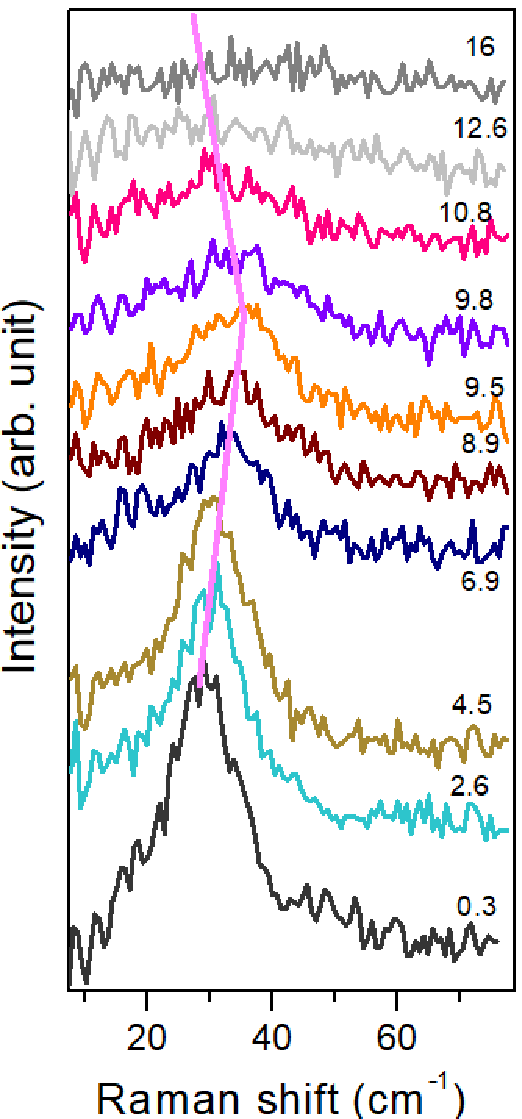}
	\includegraphics[width=0.34\linewidth]{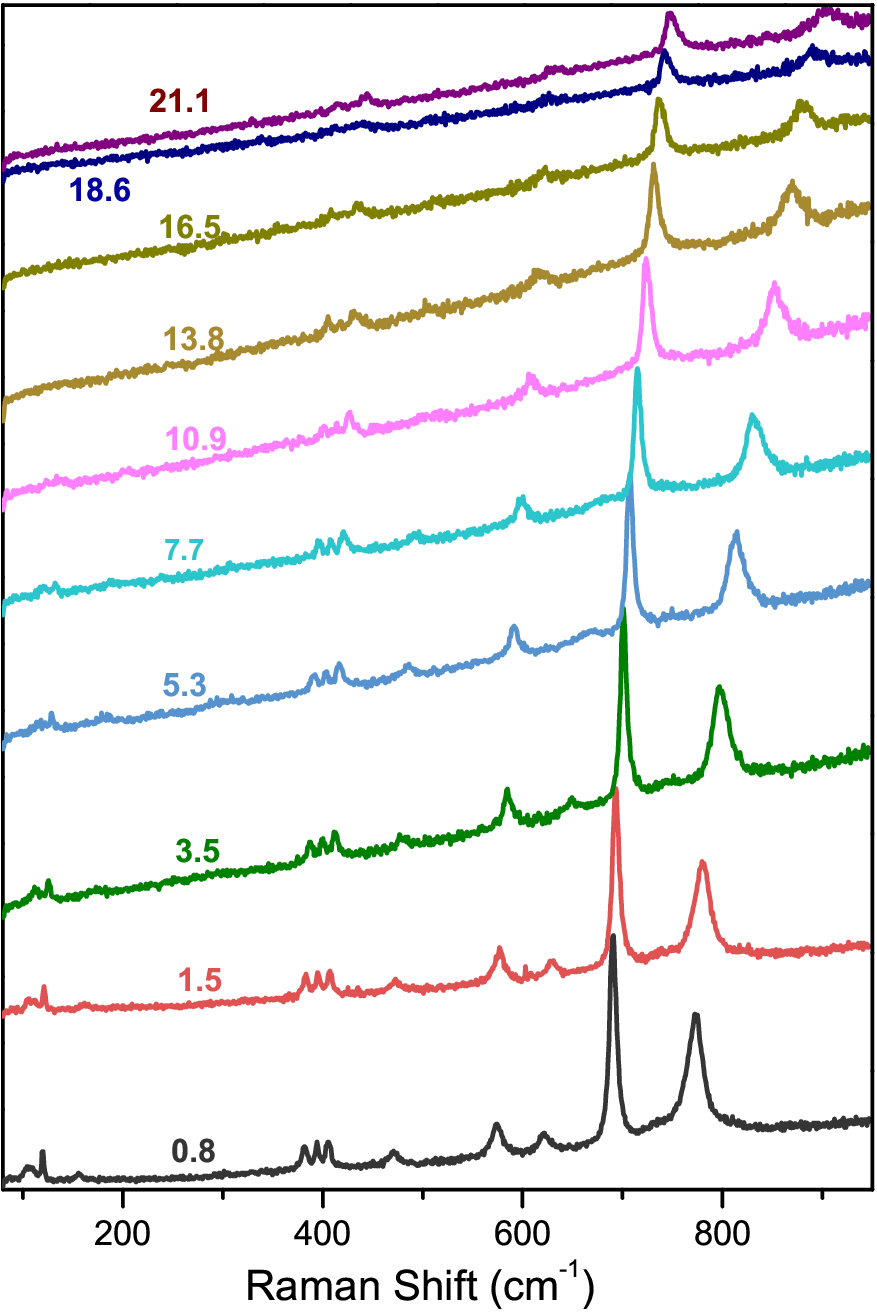}
    \includegraphics[width=0.34\linewidth]{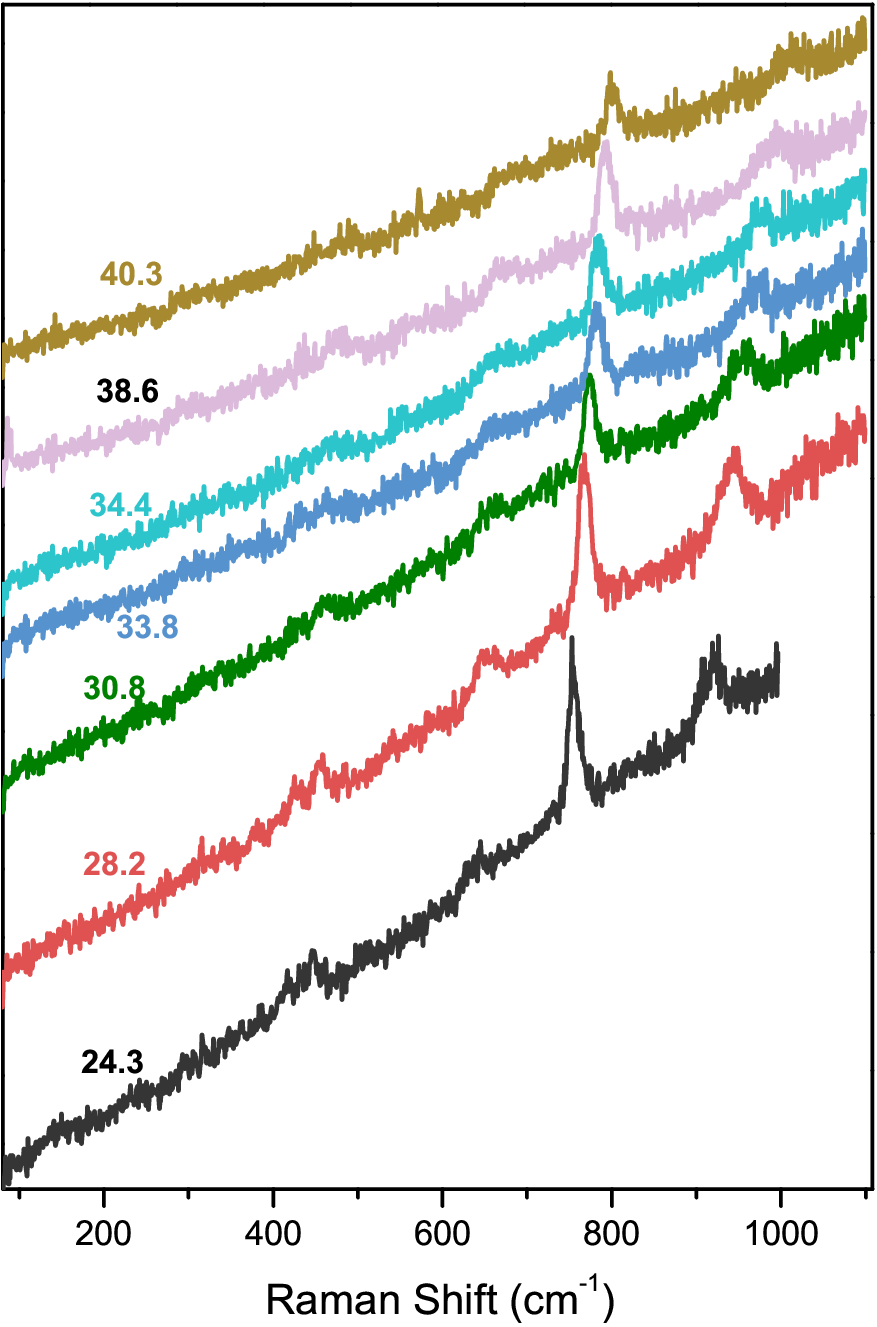}
	\caption{Pressure evolution of Raman spectra of BZTO. The pressures (in GPa) of each spectrum are indicated alongside.}
	\label{fig:AllRaman}
\end{figure}

Raman spectroscopy allows us to probe changes in the vibrational modes of the materials, offering insights into
their structural dynamics under extreme conditions. The ambient Raman spectrum matches the reported Raman spectrum at room temperature \cite{BZTOmoreiraPRM}. According to group theoretical analysis, 16 modes (7A$_{1g}+9$E$_{g}$) are Raman active for $R\bar{3}m$ structure. However, our Raman spectroscopic measurements reveal 15 Raman active modes at ambient conditions. The Raman modes are fitted to Lorentzian profiles and shown in FIG.~\ref{fig:AmbRaman}. The peak centers along with their origin are tabulated in Table~\ref{table:1}.
\begin{table}[ht!]
	\addtolength{\tabcolsep}{-3pt}
	
	\begin{center}
		\begin{tabular}{ |c|c|c|c|c|c|c|} 
	\hline
	&  &  \multicolumn{4}{c|}{Raman shift (cm$^{-1}$)} & \\ \cline{3-6}
	Mode & Sym. & \multicolumn{2}{c|}{This work} & \multicolumn{2}{c|} {Previous work} &  Contribution\\ \cline{3-6}
	& &Expt. & DFT &\cite{BZTOSurajitSaha} &  \cite{BZTOmoreiraPRM} &  \\
	\hline
	N$_1$ & E$_g$ & 28.2 & -67.6 & 31 & 28.8 & Translation of Ba, Zn, Rotation of TeO$_6$\\
	N$_2$ & A$_{1g}$ & 89 & 76 & 87 & 86 &Rotation of TeO$_6$ and ZnO$_6$\\
	N$_3$ & E$_g$ & 101.3 & 90.5 & 103 & 104.3 & Translation of Ba,Zn Rotation of ZnO$_6$\\
	N$_4$ & A$_{1g}$ & 107.9 & 101.7 & 110 & 109.7 & Translation of Ba,Zn,O \\
	N$_5$ & E$_g$ & 117.8 & 115.4 & 120 & 121.4 & Translation of Ba, Zn, Rotaion of TeO$_6$ \\
	N$_6$ & E$_g$ & 151.4 & 133.6 & 153 & 153.7 & Translation of Ba, Rotaion of TeO$_6$\\
	N$_7$ & E$_g$ & 379.7 & 362.4 & 382 & 382.7 & Scissoring of O\\
	N$_8$ & E$_g$ & 392.5 & 372.6 & 394 & 395.5 & Scissoring of O\\
	N$_9$ & A$_{1g}$ & 403.2 & -- & 405 & 406.2 & \\
	N$_{10}$ & A$_{1g}$ & 469.1 & 415.5 & 470 & 471.3 & Symmetric stretching of O\\
	N$_{11}$ & A$_{1g}$ & 570.4 & 556.5 & 573 & 572 & Symmetric and asymmetric stretching of O\\
	N$_{12}$ & E$_g$ & 614.7 & 601.7 & 616 & 616 & Symmetric and asymmetric stretching of O\\
	N$_{13}$ & E$_g$ & 688.3 & 687.3 & 689 & 690.8 & Symmetric stretching of O\\
	N$_{14}$ & A$_{1g}$ & 739.6 & 736 & 736 & 756 & Symmetric stretching of O\\
	N$_{15}$ & E$_g$ & 765.2 & - & 766 & 767.7 &\\
	\hline
\end{tabular}\\
\end{center}
\caption{Peak positions and assignments of BZTO Raman modes at ambient conditions of our experiment and reported literature.}
\label{table:1}
\end{table}
The evolution of Raman spectra at some selected pressure is shown in FIG.~\ref{fig:AllRaman}. The intensity of all Raman modes are found to decrease with pressure. 

First, we would like to focus on the behavior of the N$_1$ mode (E$_g$ symmetry) under the application of pressure. The behaviour of peak center of N$_1$ mode (28.2~cm$^{-1}$) with increasing pressure is shown in Supplementary FIG.~S2~\cite{supp}. When pressure is applied to a material, the interatomic distances decrease, which increases the bonding strength. As a result, the frequency of vibrational modes, including the Raman-active mode increases as the mode requires more energy to be excited. With an increase in pressure, the N$_1$ mode shows typical hardening behavior up to 9.1~GPa followed by an unusual softening. As the intensity of this mode is very low and with increasing pressure the intensity decreases further, we could not detect the mode above 12.6~GPa.
Generally, ``soft mode" mechanism can account for structural phase transitions in a variety of situations. The theory of soft mode phase transitions was first developed in the context of ferroelectric crystals where the ``soft mode" was a transverse optic mode of long wavelength. The temperature dependence of this mode was developed by Cochran as \cite{cochran}

\begin{equation} \omega(T) = A \sqrt{T - T_C}.\end{equation}

We find that the mode at 28.2~cm$^{-1}$ has strong anharmonicity and shows anomalous softening as the pressure is increased. In particular it does not follow the expected linear dependence on pressure but follows the modified Cochran-type relation \cite{cochran_pressure}

\begin{equation} \omega(P) = A \sqrt{P_C - P}.\end{equation}

\begin{figure}
	\centering
		\hspace{-6.5cm}(a) \hspace{0.45\linewidth} (b) \\
	\includegraphics[width=0.46\linewidth]{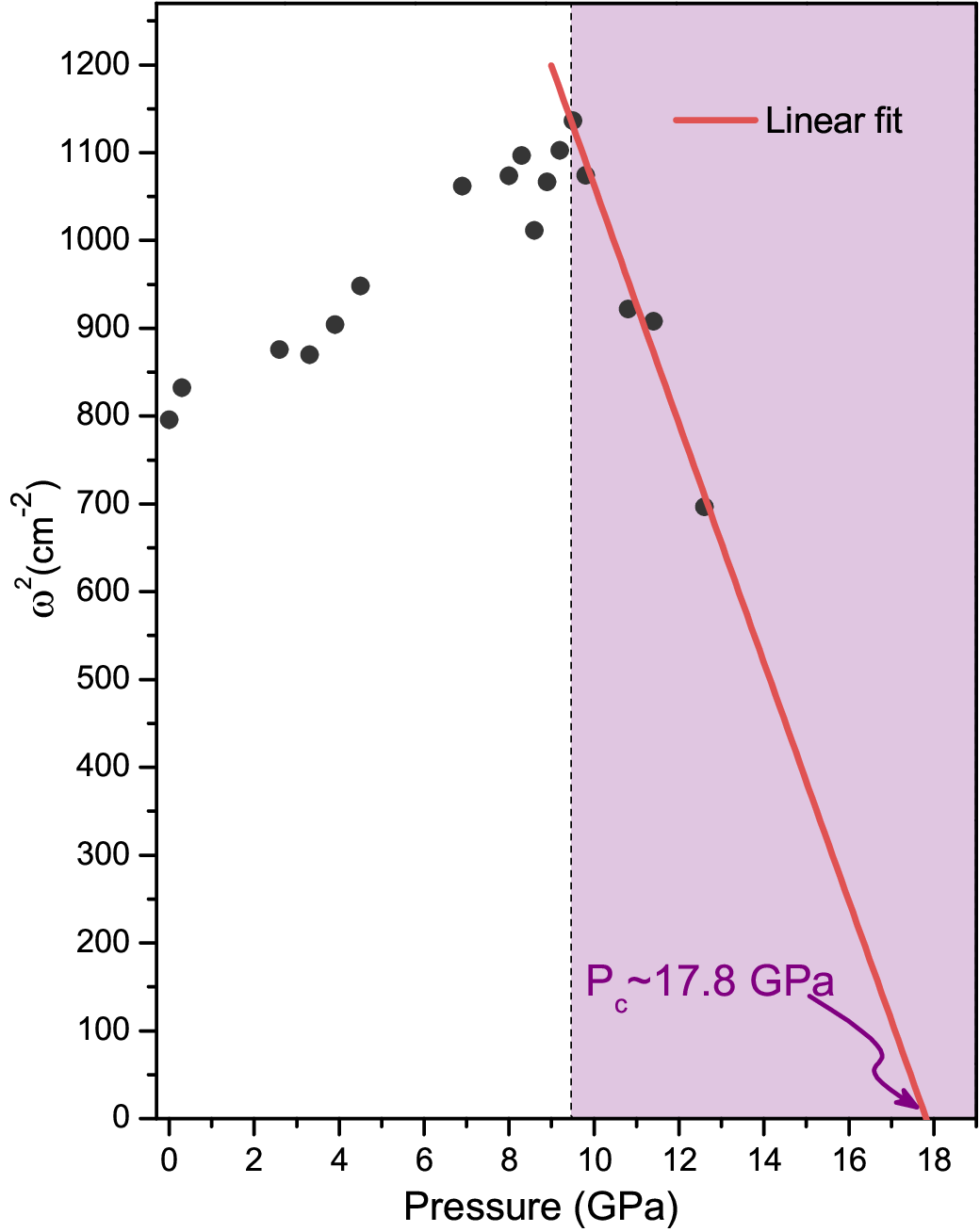}
	\includegraphics[width=0.460\linewidth]{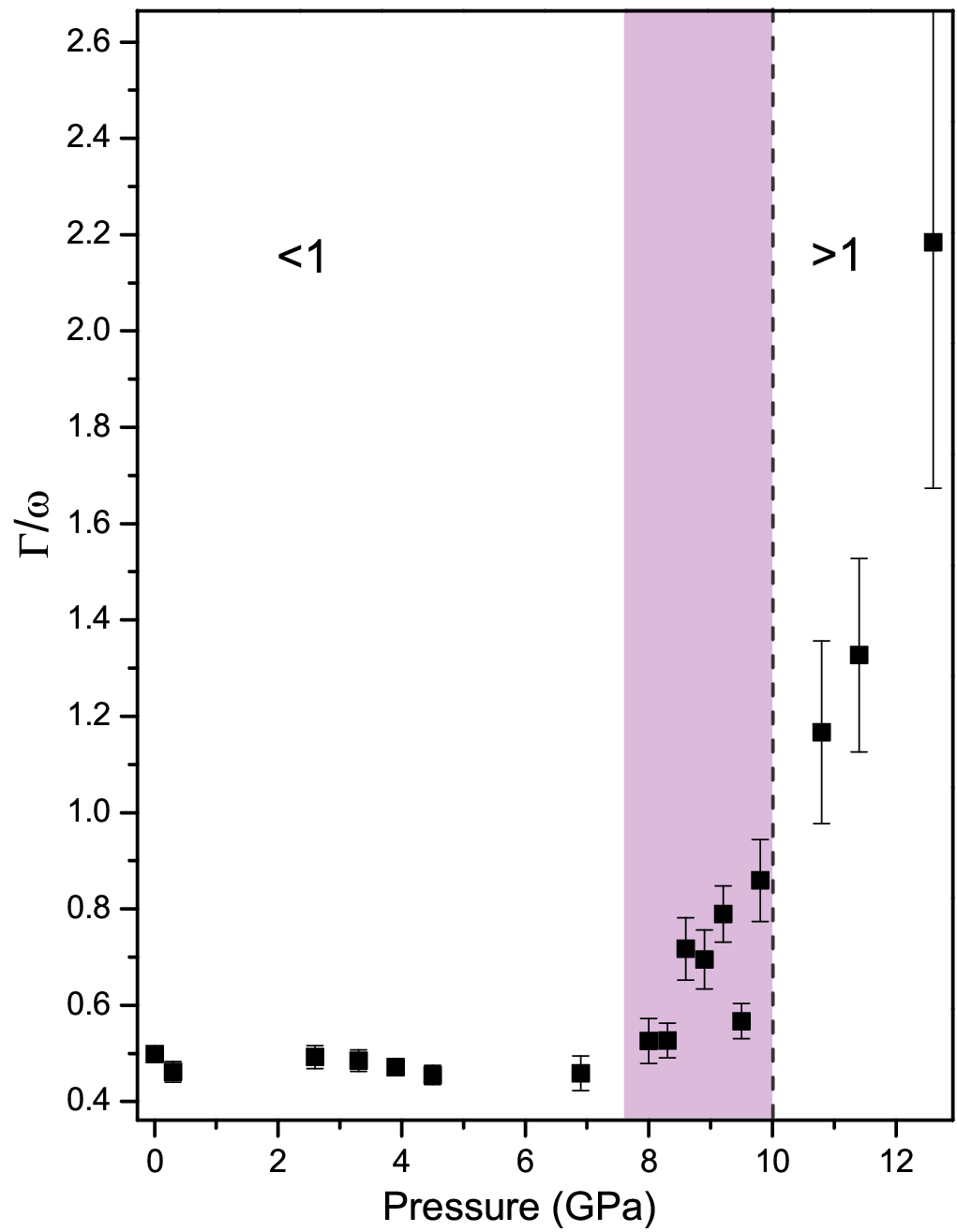}
	\caption{(a) Pressure evolution of square of soft mode frequency. (b) Pressure evolution of $\frac{\Gamma}{\omega}$ of the soft mode. The dashed line separates the under-damped and over-damped regions.}
	\label{fig:SoftMode}
\end{figure}

The mode becomes faint at 12~GPa and cannot be distinguished from the noise in the Raman spectra. Remarkably, the fitting of the mode with the modified Cochran-type relation is excellent, and on extrapolating it beyond the last discernible pressure point, it intersects the pressure axis at 17.8 GPa (see FIG.~\ref{fig:SoftMode}(a)). This indicates that the soft mode has become unstable and there is the possibility of a structural phase transition beyond that pressure. Using XRD analysis, we can verify this prediction. If the full-width half maxima of a soft mode is smaller than the frequency of that mode then the mode is underdamped. To check that, we have plotted $\frac{\Gamma}{\omega}$ versus pressure as shown in FIG.~\ref{fig:SoftMode}(b), which shows that $\frac{\Gamma}{\omega}<1$ up to 10~GPa i.e. up to 10~GPa it is underdamped however the damping is almost constant up to $\sim $7.5~GPa and increases with pressure afterwards. Upon compressing above 10~GPa it becomes $>1$. This implies that above 10~GPa the underdamped mode becomes overdamped and the damping increases with pressure in the softening region. The overdamping in the soft mode leads to continuous structural phase transition due to strong anharmonicity in the lattice vibrations. This is apparent in the observed increase in FWHM of Raman modes above 10~GPa.

\begin{figure}
	
	\hspace{-14.5cm}(a)\\ 
	\includegraphics[width=0.85\linewidth]{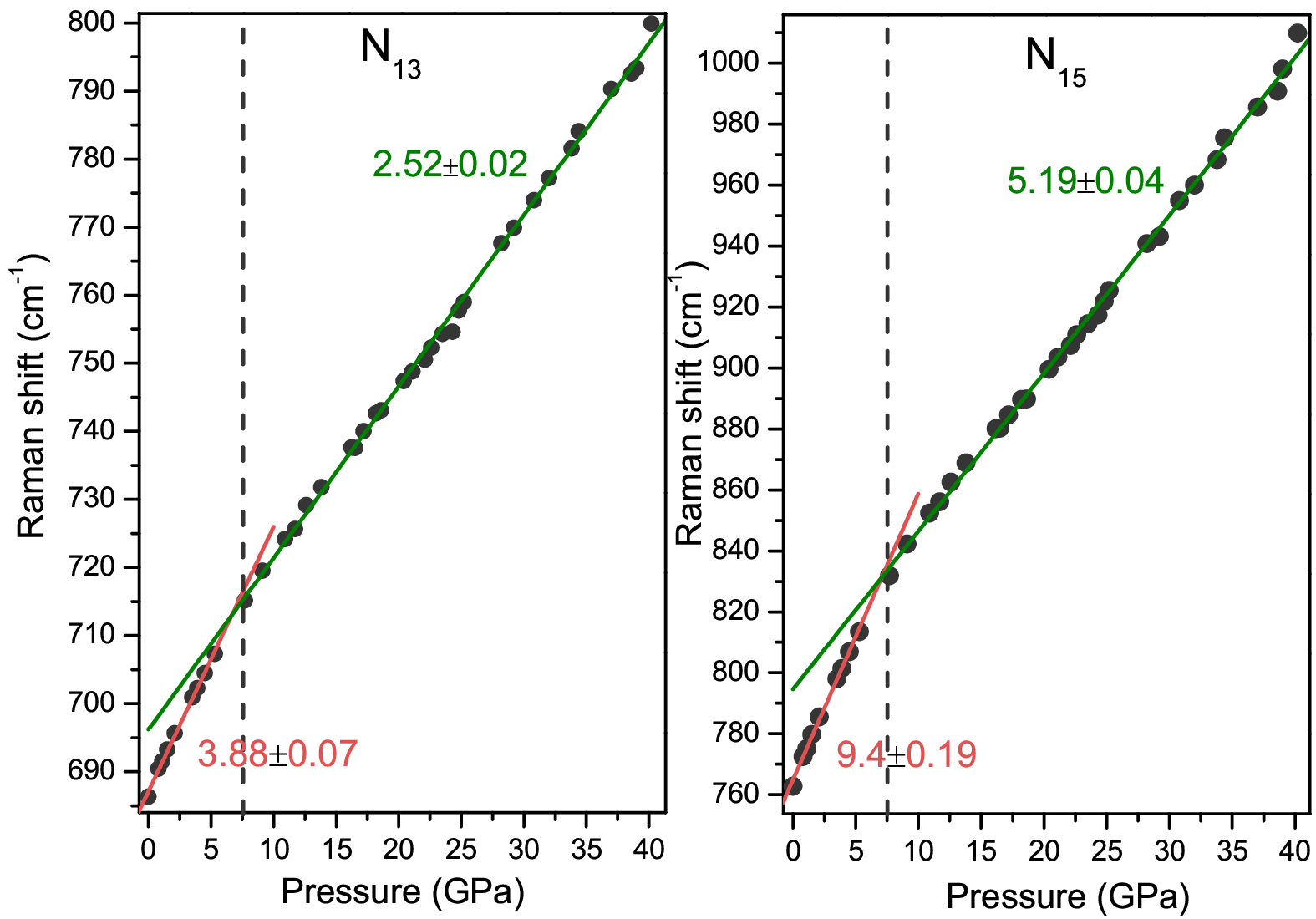}\\ \hspace{-14.5cm}(b)\\
	\includegraphics[width=0.85\linewidth]{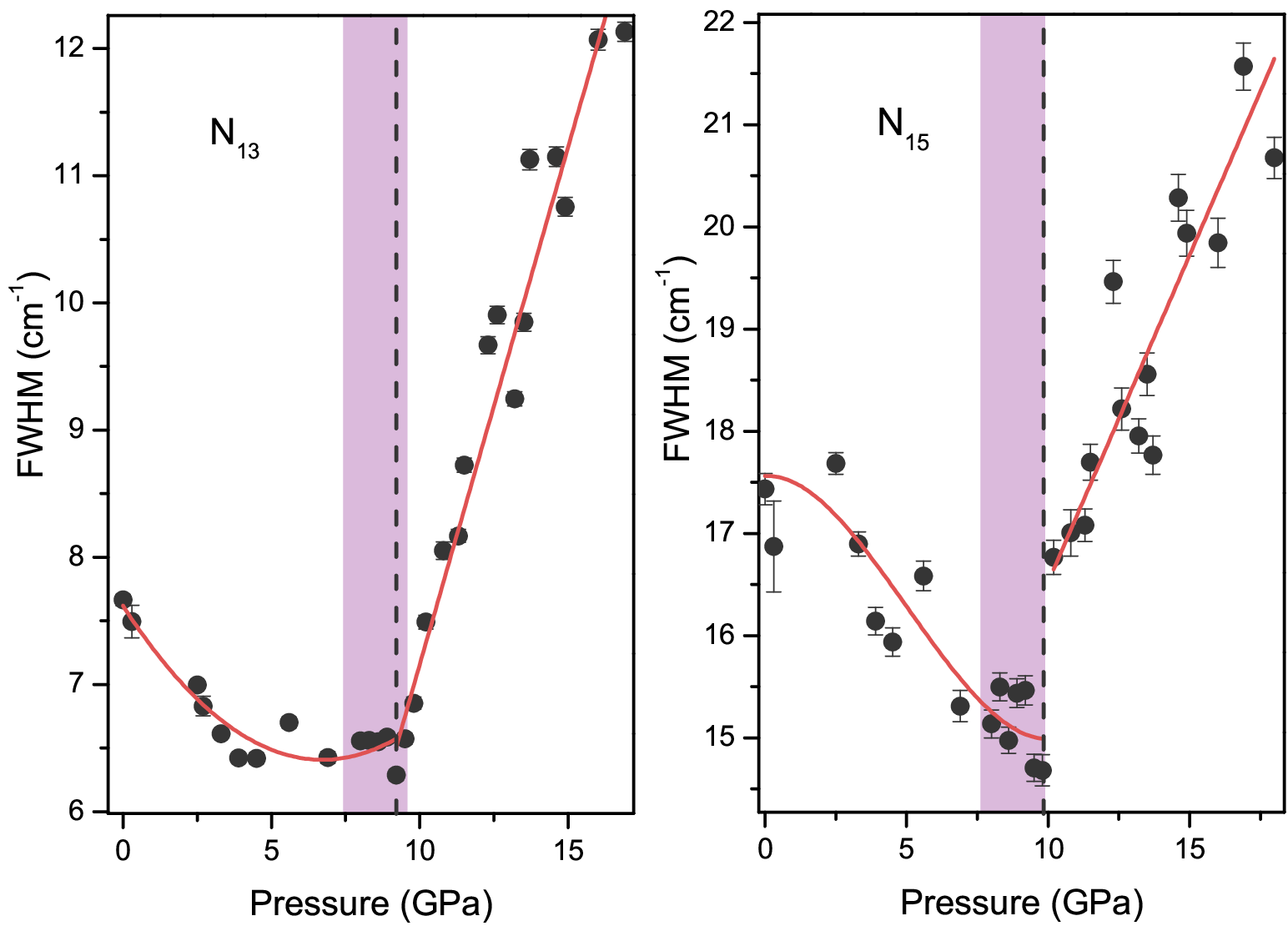}
		\caption{(a) Pressure evolution of peak positions of N$_{13}$ and N$_{15}$ Raman modes. The evolution of peak positions with pressure is fitted using linear fit. The slope of each linear fit is indicated alongside. (b) Pressure evolution of FWHM of N$_{13}$, N$_{15}$ Raman modes.}
		\label{fig:PeakCenter}
\end{figure}

 The prominent Raman modes, N$_{13}$ (688.3~cm$^{-1}$) and N$_{15}$ (765.2~cm$^{-1}$) are plotted against pressure in FIG.~\ref{fig:PeakCenter}(a), which shows a slope change around 7.5~GPa. The values of the slopes of each mode below 7.5~GPa are greater than those above 7.5~GPa. The slope values are mentioned alongside each linear fitting. Similar decrease in slope of the Raman mode frequencies are observed in all other Raman modes, some of which (N$_7$ (379.7~cm$^{-1}$), N$_8$ (392.5~cm$^{-1}$), N$_9$ (403.2~cm$^{-1}$), N$_{11}$ (570.4~cm$^{-1}$)) are shown in the Supplementary Information FIG.~S3~\cite{supp}.
 The N$_1$ mode also shows a slope change of around 8~GPa before the softening starts around 10~GPa. This may indicate that the structure is well resistant to pressure up to 7.5~GPa and starts to lose its stability in the pressure region of 7.5 to 10~GPa causing all the Raman mode frequencies to change its slope with respect to pressure. The FWHM of the Raman mode N$_{15}$ decreases up to 10~GPa and then suddenly increases from 10~GPa (see FIG.~\ref{fig:PeakCenter}(b)). The FWHM of N$_{13}$ is almost constant between the pressure range of 7.5~GPa to 10~GPa before starting to increase with pressure. Increase in FWHM with pressure indicates that structure is distorting as the pressure increases above 10~GPa. High-pressure XRD analysis can give further insight into this.

\subsection{High Pressure XRD study}
\begin{figure}
	\centering
	\includegraphics[width=0.65\linewidth]{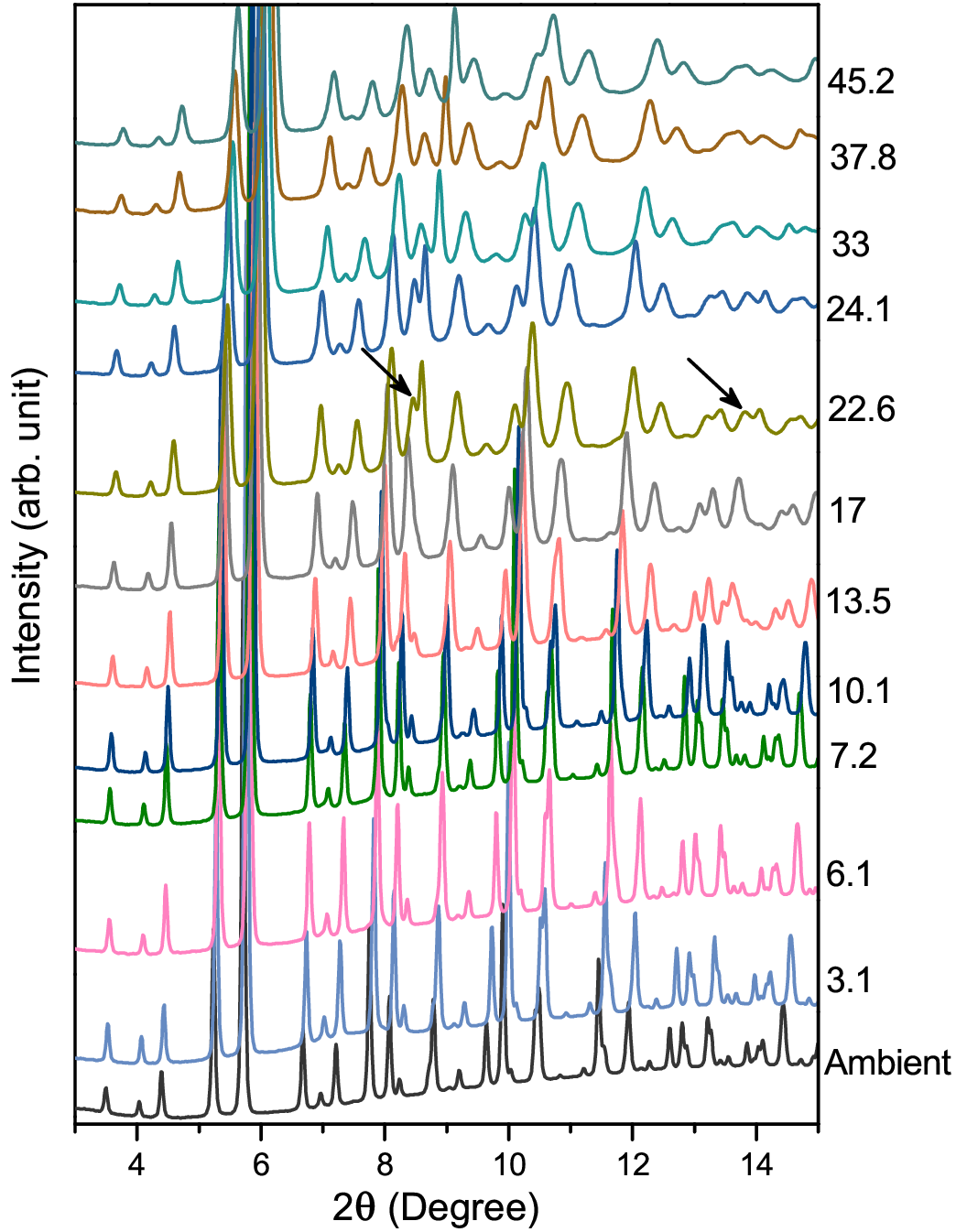}
	\caption{XRD pattern of BZTO taken at different pressures. Pressures (in GPa) are indicated alongside the respective data. The black arrows indicate the appearance of new peaks.}
	\label{fig:AllXRD}
\end{figure}

High-pressure synchrotron XRD measurements were carried out up to 45~GPa.
The pressure evolution of XRD data of some selected pressure points is shown in FIG.~\ref{fig:AllXRD}. All the Bragg peaks shifted to the right side with increasing pressure. All the XRD data at and below 18~GPa can be fitted well using the parent structure. Above 18~GPa the structure can not be fitted using the parent space group due to the appearance of some new Bragg lines. Remarkably, the softening of the N$_1$ phonon mode predicts structural transition at 17.8~GPa. The black arrows indicate the new Bragg peaks which appeared at 22.6~GPa. A careful indexing leads to a monoclinic structure with space group $C2/m$ above 18~GPa. Lattice parameters obtained from refinements are $a= 9.6661(9)$~\si{\angstrom}, $b= 5.5424(2)$~\si{\angstrom}, $c= 9.7098(5)$~\si{\angstrom}, $\beta= 108.842(6)$\textdegree  ~and the corresponding unit cell volume is $492.31(3)$~\si{\angstrom}$^3$ at 22.6~GPa. 
A similar structural transition from the paraelectric trigonal to the ferroelastic monoclinic $C2/m$ phase is observed below 140 K by Badola et.al. \cite{BZTOSurajitSaha} with a large softening of low frequency $E_g$ mode and appearance of central peak in Raman spectrum.Interestingly, in the present case the phase transition to the monoclinic phase does not seem to be ferroelastic from the absence of any central peak in the Raman spectra at high pressures. It is a typical phenomenon that the low temperature phase is stabilized by application of high pressure. Increasing pressure compresses the system, leading to a denser packing, while lowering temperature reduces thermal fluctuations, also leading to compaction. Although pressure and temperature influence materials differently at the microscopic level, their effects on the free energy landscape often mirror each other, guiding the system toward similar phases.
To carry out the Rietveld refinement, the atomic positions were taken from a sister compound, Ba$_2$CuTeO$_6$ which has a monoclinic $C2/m$ space group \cite{BCuTOSurajitSaha}. The Rietveld refinement at 24.1~GPa is shown in FIG.~\ref{fig:XRD}. The R$_p$ and R$_{wp}$ values are 0.7\% and 0.99\% respectively, which indicates an excellent fit. The refined relative atomic positions at 24.1~GPa are given in Table.~\ref{table:3}.
\begin{table}[ht!] 
\begin{center}
\begin{tabular}{ |c|c|c|c|c|c| } 
 \hline
 Atom & Wyckoff position & x/a & y/b & z/c\\
 \hline
 Ba(1) & 4i & 0.13343(1) & 0.0 & 0.38193(1)\\
 Ba(2) & 4i & 0.294459 & 0.25 & 0.83889(2)\\
 Zn & 4i & 0.9288(2) & 0.5 & 0.2318(3)\\
 Te(1) & 2a & 0.0 & 0.0 & 0.0\\
 Te(2) & 2b & 0.5 & 0.0 & 0.5\\
 O(1) & 4i & 0.1328 & 0.5 & 0.3998\\
 O(2) & 8j & -0.1046 & 0.7283 & 0.3687\\
 O(3) & 4i & 0.3177 & 0.5 & 0.8754\\
 O(4) & 8j & 0.0608 & 0.6603 & 0.8999\\
 \hline
\end{tabular}\\
\end{center}
 \caption{Relative atomic positions at 24.1~GPa after Rietveld refinement. The unit cell lattice parameters are: $a= 9.753(3)$~\si{\angstrom}, $b= 5.522(2)$~\si{\angstrom}, $c= 9.662(4)$~\si{\angstrom}, $\beta= 109.79(4)$\textdegree and corresponding volume is 489.6(2)~\si{\angstrom}$^3$.}
\label{table:3}
\end{table}

\begin{figure}
	\centering
	\hspace{-5cm}(a) \hspace{0.4\linewidth} (b) \\
	\includegraphics[width=0.41\linewidth]{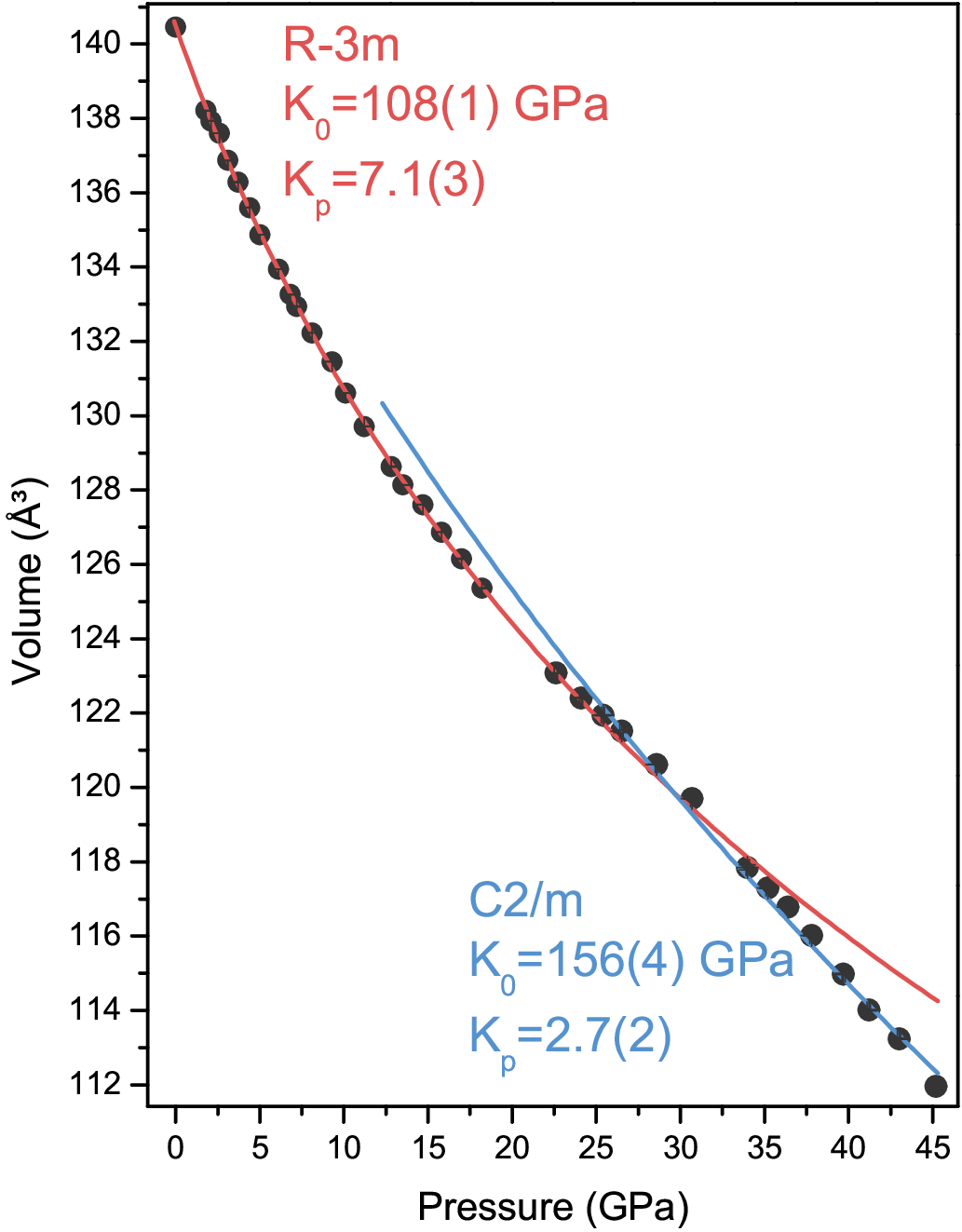}
	\includegraphics[width=0.385\linewidth]{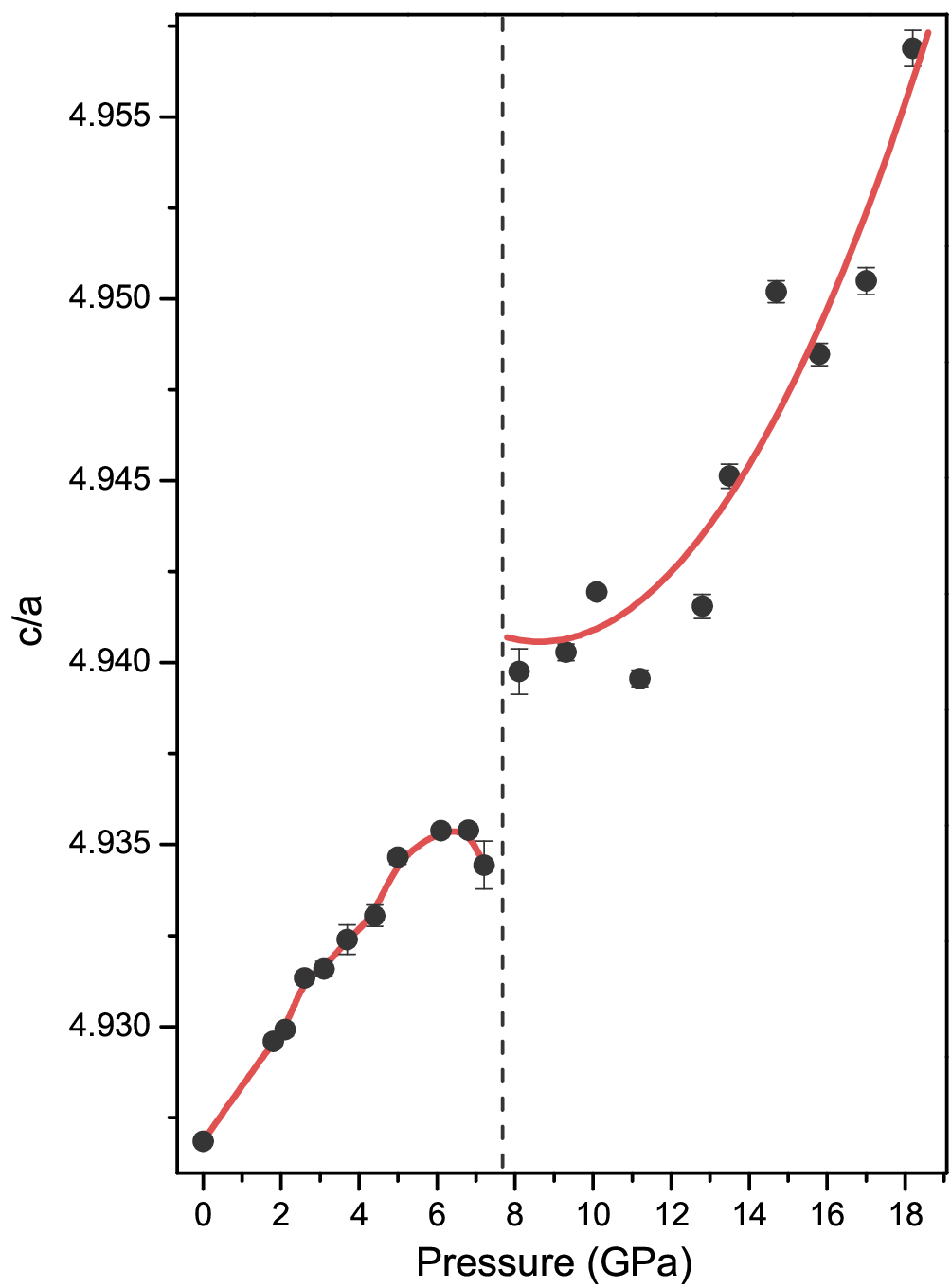}
	\caption{(a) 3$^{rd}$ order Birch-Murnaghan equation of state fit to volume versus pressure data. Solid black circles represent the experimental unit cell volume data normalized with respect to formula unit, red and blue line represents the EoS fit to it for rhombohedral and monoclinic phase respectively. (b) $c/a$ ratio with respect to pressure. }
	\label{fig:LattParams}
\end{figure}

To see the pressure variation of volume, we first normalized the unit cell volume with respect to formula unit and plotted it against pressure in FIG.~\ref{fig:LattParams}(a). The volume data of both the phases are fitted using the 3$^{rd}$ order Birch-Murnaghan (BM) equation of state (EoS) \cite{birch1947,murnaghan1944compressibility}. The obtained bulk modului are 108(1)~GPa and 156(4)~GPa and the first derivatives of bulk modulus with pressure are 7.1(3) and 2.7(2) for $R\bar{3}m$ and $C2/m$ phases respectively.

\begin{figure}
	\centering
	\includegraphics[width=0.95\linewidth]{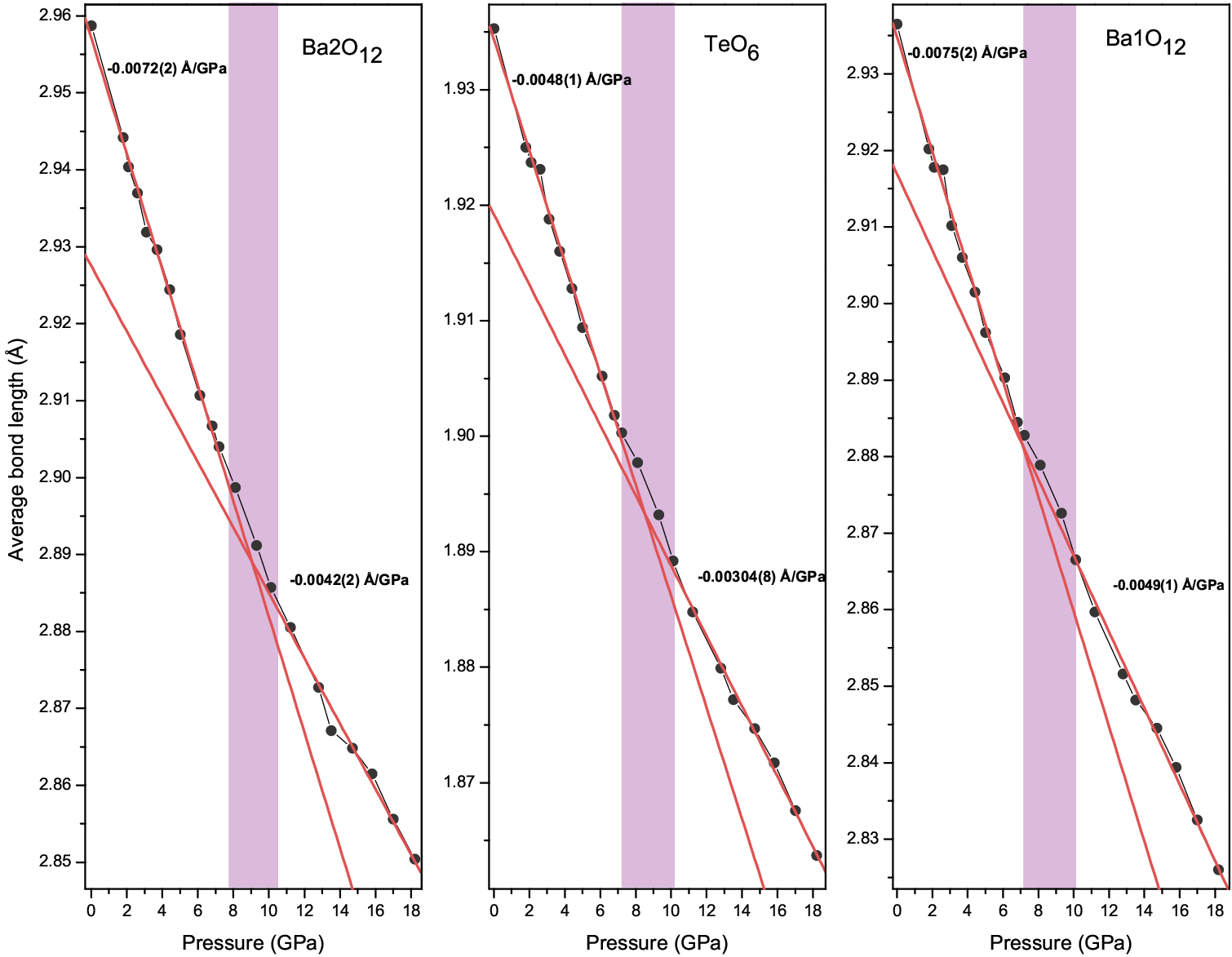}
	\caption{Pressure variation of average bond lengths of BaO$_{12}$ polyhedra and TeO$_6$ octahedra.}
	\label{fig:Bondlength}
\end{figure}

Now we try to understand the rhombohedral phase to find out the physics behind the anomalous behavior of Raman modes in the pressure region between 7.5~GPa to 10~GPa as discussed in the earlier section. The c/a ratio, which is indicative of unit cell distortion, is plotted with pressure in FIG.7(b). Interestingly, it shows a linear increase with pressure then shows a dip around 7.5~GPa followed by a sudden jump and then increases at a higher rate with pressure afterwards as shown in FIG.~\ref{fig:LattParams}(b). This shows an anisotropic compression of the unit cell with the a-axis being more compressible compared to the c-axis as shown in the supplementary figure, FIG.~S5~\cite{supp}. An increase in trigonality with pressure signifies an increase in instability in the crystal structure. FIG.~\ref{fig:LattParams}(b) depicts that, the structure tries to maintain its stability in the vicinity of 7.5~GPa, however, it becomes more and more unstable as the pressure increases beyond 7.5~GPa. Next, we have plotted the average bond length of TeO$_6$ octahedra and two types of BaO$_{12}$ polyhedra in FIG.~\ref{fig:Bondlength}. The average bond lengths of Ba(1)O$_{12}$ and Ba(2)O$_{12}$ polyhedra decrease with pressure at a rate of -0.0075(2)~\AA/GPa and -0.0072(2)~\AA/GPa respectively below 7.5~GPa and at a rate of -0.0049(1)~\AA/GPa and -0.0042(2)~\AA/GPa respectively above 10~GPa. Similarly, the rate of change of the average Te--O bond length of TeO$_6$ octahedra with pressure decreases from -0.0048(1)~\AA/GPa (below 7.5~GPa ) to -0.00304(8)~\AA/GPa (above 10~GPa). The variation of average bond lengths with respect to pressure shows that the bond lengths are more compressible below 7.5~GPa and above 10~GPa they become stiffer. It is well known that any subtle changes in the structural dynamics can be captured by Raman spectroscopic analysis. This has indeed been seen in the Raman analysis (FIG.~\ref{fig:PeakCenter}), that is the increase in the Raman mode frequencies are notably way faster up to 7.5~GPa than the rate of change afterwards. We have also discussed the pressure variation of bond angle variance in supplementary information \cite{supp}, as it is correlated with the structural stability. So it can be inferred that the anomalous increase in bond angle variance and structural trigonality are the key causes of making the sample unstable above 7.5~GPa and the same is reflected in the pressure evoluation of Raman modes.


\section{Computational study}
\begin{figure}
	\centering
	\includegraphics[width=0.450\linewidth]{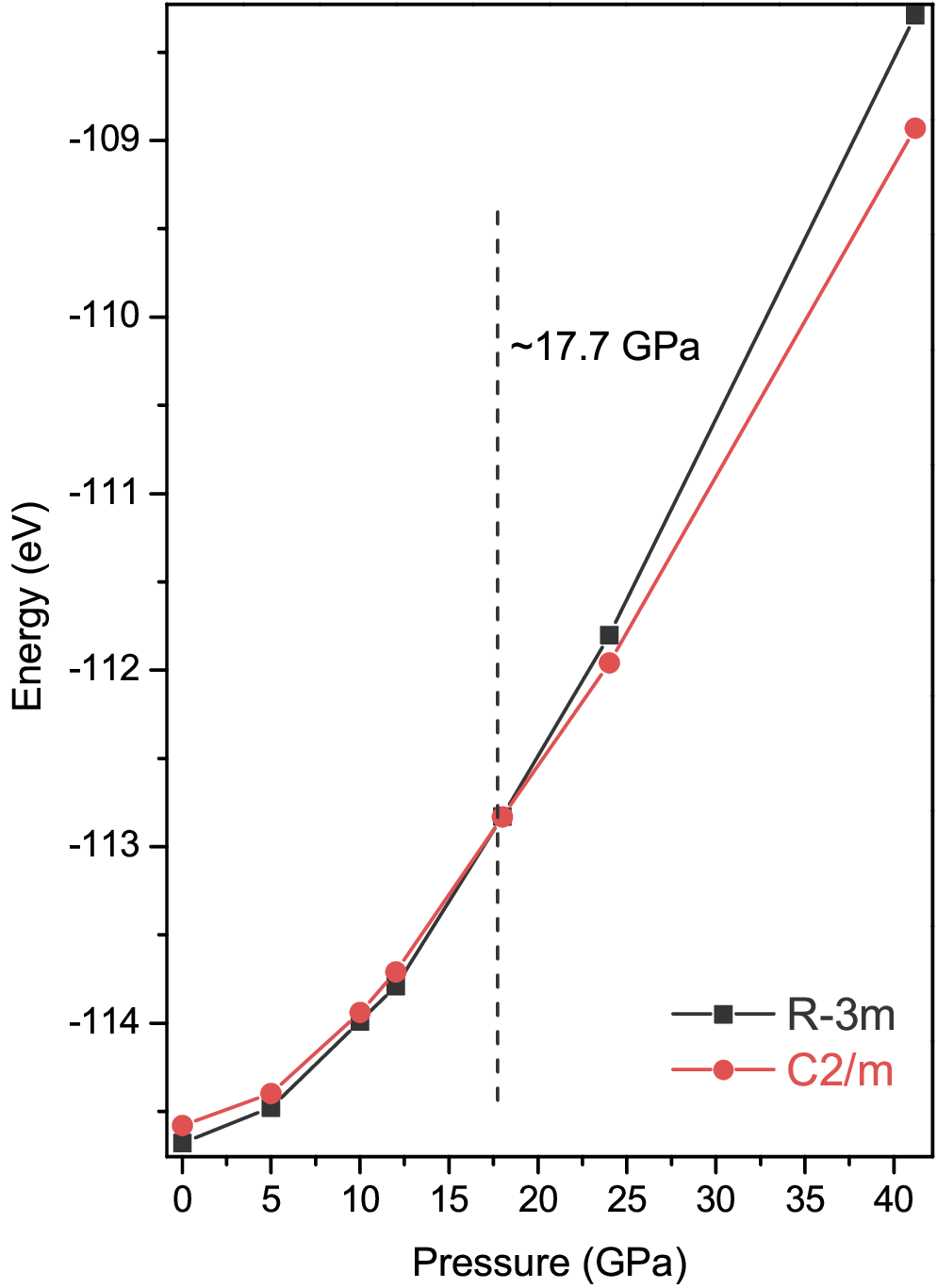}
	\caption{Comparison of the free energy for both the phases as a function of pressure. Dashed line separates the $R\bar{3}m$ and $C2/m$ phases. }
	\label{fig:TotalEnergy}
\end{figure}

\begin{figure}
    \centering
    \includegraphics[width=0.8\textwidth,height=1\textheight,keepaspectratio]{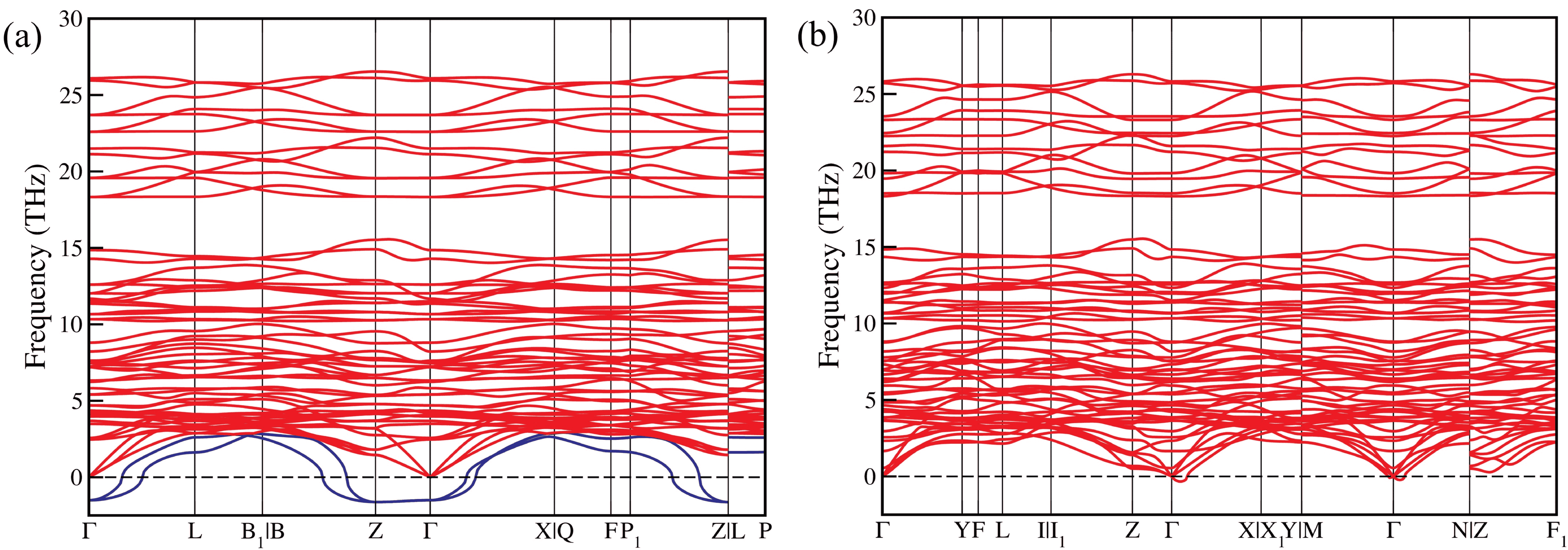}
    \caption{\label{fig:Phonon}Phonon dispersion curves of (a) trigonal (space group
$R\bar{3}m$) phase and (b) monoclinic (space group $C2/m$) phase for
Ba$_{2}$ZnTeO$_{6}$ crystal structure at 24~GPa calculated using DFPT method.}
\end{figure}

To validate the experimental results, we have carried out DFT-based simulations and assessed the dynamical (vibrational) stability as well as electronic properties of the trigonal (space group $R\bar{3}m$) and monoclinic (space group $C2/m$) phase for Ba$_{2}$ZnTeO$_{6}$ crystal structure. Initially, we fully relaxed the atomic coordinates, volume, and lattice parameters of these phases using the PBE xc functional. As a result, we obtained lattice parameters of $a = b = 5.9153~\si{\angstrom}$, $c = 29.1443 ~\si{\angstrom}$ and a volume of 883.1568~\si{\angstrom}$^{3}$ for the R$\bar{3}$m crystal structure, which are in good agreement with our measured parameters obtained by XRD. Simultaneously, the calculated lattice parameters for the $C2/m$ crystal structure at 24~GPa are $a = 9.6721~\si{\angstrom}$, $b = 5.5927~\si{\angstrom}$, $c = 9.8295~\si{\angstrom}$, and $\beta = 106.957${\textdegree}, with a corresponding unit cell volume of $500.620~\si{\angstrom}^{3}$. Pressure variation of lattice parameters of both the phases as a function of pressure are tabulated in supplementary material~\cite{supp}. Though the structural transition is evident from experimental work, to further explore the phase transition using DFT, we have calculated the total energy of both the systems (trigonal and monoclinic) and plotted them as a function of pressure in FIG.~\ref{fig:TotalEnergy}. The plot shows the $R\bar{3}m$ phase has lower energy compared to the $C2/m$ phase from ambient up to around 18~GPa. At and above that the monoclinic phase processes the lower energy, which shows that the $C2/m$ crystal structure becomes more stable at and above $\sim$18~GPa pressure region and drives the structural transition. Interestingly, we identified two unstable modes for the trigonal phase at 24~GPa, as evidenced by the negative frequency values in the phonon dispersion curve [see FIG.~\ref{fig:Phonon}(a)]. These unstable modes are due to a doubly degenerate mode ($E_{g}$) at $\Gamma$-point. The $E_{g}$ mode involves an in-phase rotation of TeO$_{6}$ octahedra with the rotation axis lying within the xy-plane (see Table I). The unstable $E_{g}$ mode leads to a structural phase transition, which is in complete agreement with the experimental findings discussed above. From the phonon dispersion curve at 24~GPa of the monoclinic phase [see FIG.~\ref{fig:Phonon}(b)], it is evident that all vibrational modes exhibit positive frequency values, indicating that this monoclinic phase is vibrationally stable. Thus, our theoretical results corroborate the experimental observations, indicating that the Ba$_{2}$ZnTeO$_{6}$ crystal system undergoes a structural phase transition when subjected to higher pressure. 

In addition to the vibrational stability, we also carried out atom-projected partial density of states (pDOS) and electronic band structure for both phases using the GGA-PBE xc functional. Our results reveal that both phases exhibit semiconducting behavior having indirect bandgap. It is a well-known fact that the standard DFT calculations with GGA underestimates the actual bandgap \cite{HSE06}. However, it captures the correct nature of pressure evolution of the bandgap. Sophisticated functionals, such as HSE06, HLE16, and mBJ can predict more accurate bandgap \cite{HSE06}. But here we avoided band-structure calculations using these functionals except for ambient structure due to high computational costs. The band structures using both HSE06 xc and GGA-PBE functionals are shown in FIG.~S7.~(c) and FIG.~S8.~(c), respectively~\cite{supp}. We found the indirect bandgap is 3.62~eV using HSE06 functional which is in good agreement with our experimental results (FIG.~S7.~(b)) with the conduction band minimum (CBM) and valence band maximum (VBM) situated at the $\Gamma$-point and F-point, respectively. From FIG.~S8.~(c), it can be observed that the trigonal phase exhibits an indirect bandgap of 2.045~eV using PBE~\cite{supp}. The calculated bandgap value using GGA-PBE xc functionals is underestimated by 43\% from the calculated bandgap value using a sophisticated functional, HSE06.

\begin{figure}
    \centering
    \hspace{-5cm}(a) \hspace{0.34\linewidth}(b)\\
    \includegraphics[width=0.38\linewidth]{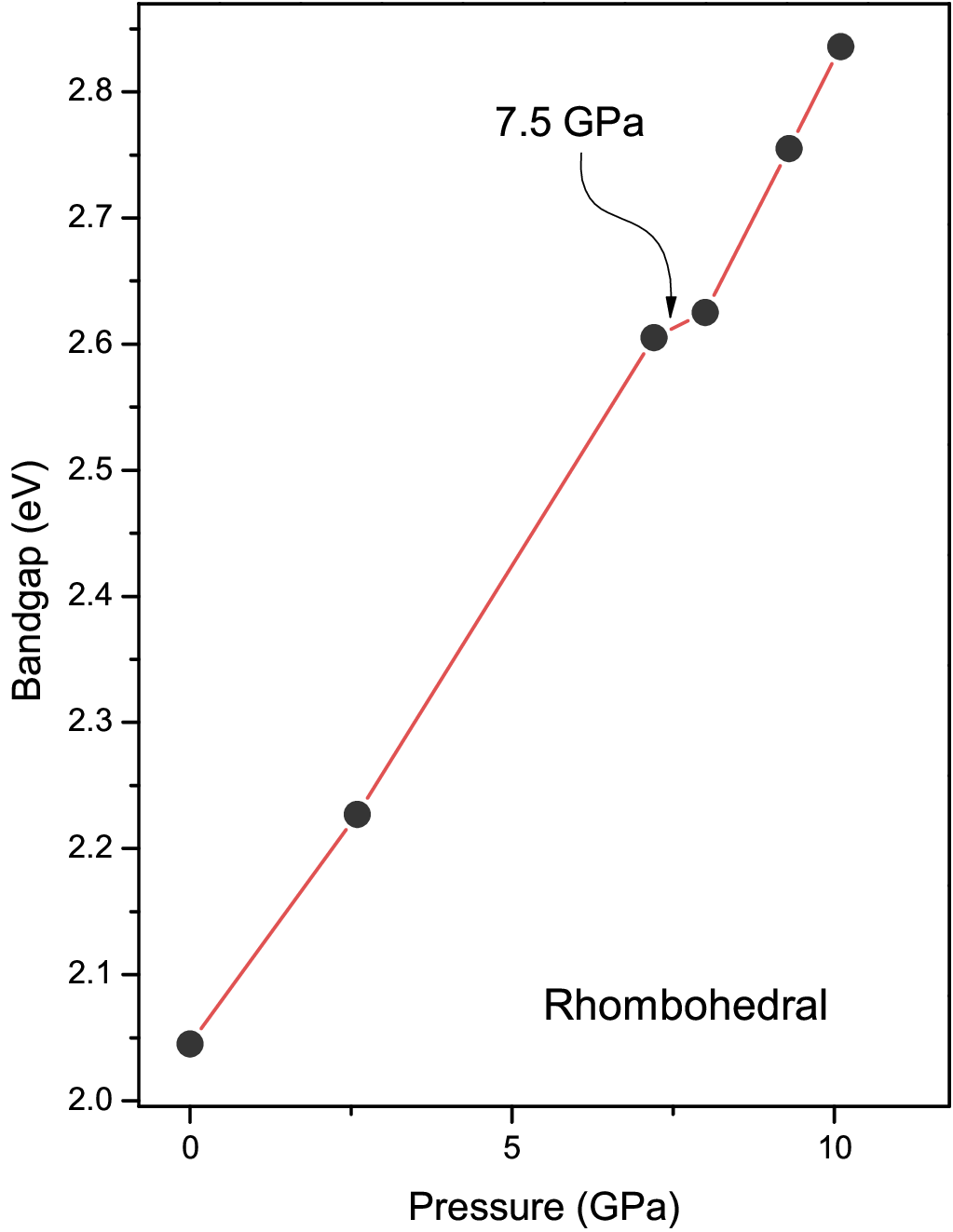}
    \includegraphics[width=0.38\linewidth]{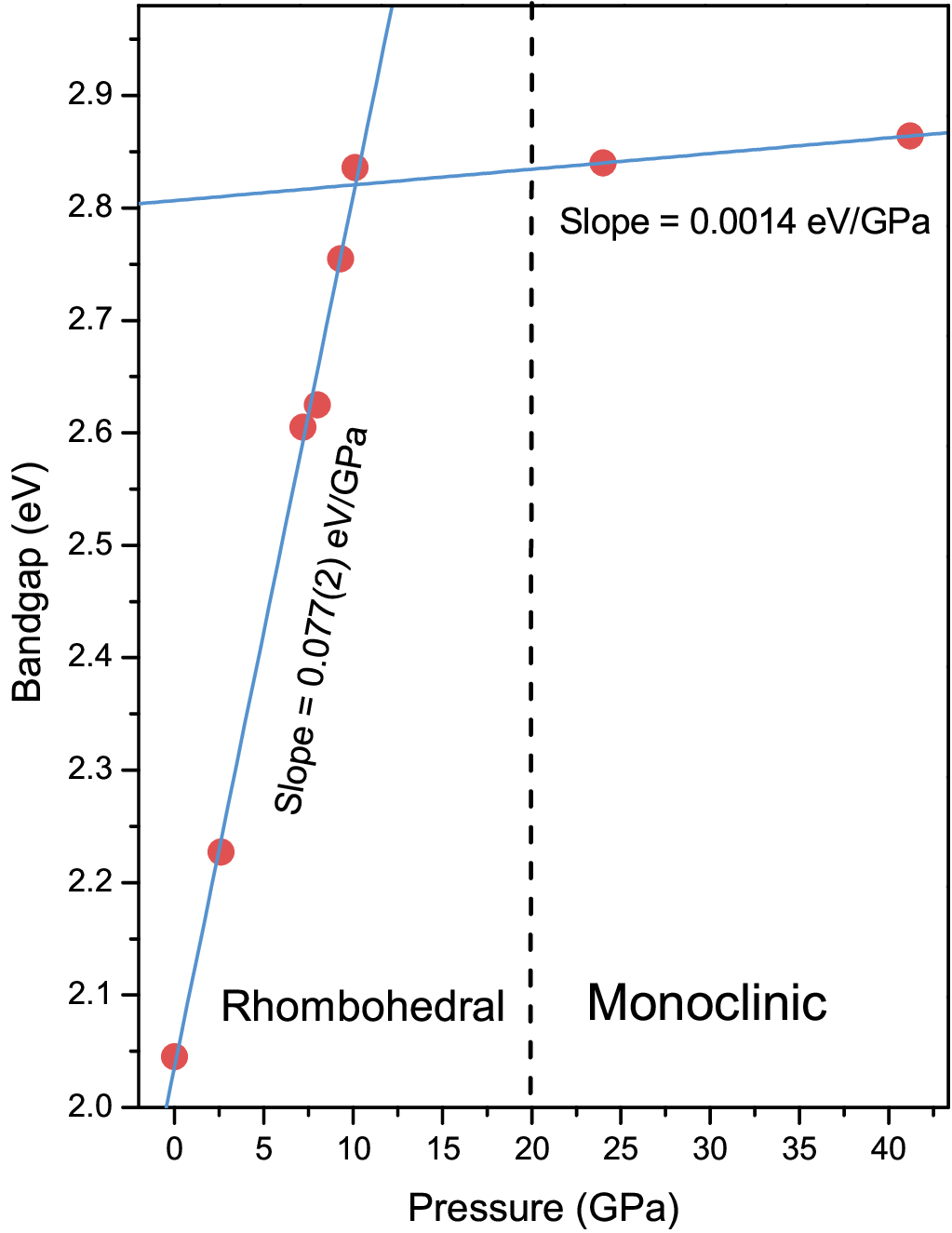}
    \caption{Variation of calculated band gap with pressure (a) for $R\bar{3}m$ phase only, (b) for both the phases. The vertical dashed line is drawn at the boundary of $R\bar{3}m$ and $C2/m$ phases of Ba$_{2}$ZnTeO$_{6}$.}
    \label{Bandgap}
\end{figure}

On the other hand, the monoclinic phase has an indirect bandgap of 2.84~eV, with the CBM and VBM positioned at the $\Gamma$-point and M-point, respectively [see FIG.~S8.~(d)] at 24~GPa~\cite{supp}. In FIG.~\ref{Bandgap} (b), the variation of bandgap with increasing pressure is shown up to 40.3~GPa. As pressure increases the bandgap also increases in the whole pressure region of rhombohedral phase. The average rate of change of bandgap with pressure is 0.077~eV/GPa in the rhombohedral phase while in the monoclinic phase, it becomes 0.0014~eV/GPa. The change in the slope of the bandgap with respect to pressure decreases significantly and it is attributed to the transition from a high-symmetry phase ($R\bar{3}m$) to a low-symmetry phase ($C2/m$). Now if we closely observe the $R\bar{3}m$ phase only [see FIG.~\ref{Bandgap}(a)], we can see the monotonous increase of bandgap with pressure interrupted around 7.5~GPa by a shift of bandgap value. This shift may be correlated to the observed dip in c/a value around 7.5~GPa followed by a sudden jump with increasing pressure. Numerous studies have been documented on the influence of ratio of lattice parameters on band structure, highlighting the intricate relationships among structural and electronic properties \cite{DebabrataSamantac/aratio,CsPbBr3_c/aratio}. For example, with increasing the ratio of lattice constants  bandgap increases with pressure in CsPbBr$_3$ perovskite \cite{CsPbBr3_c/aratio}. This indicates there is a correlation between lattice and electronic degrees of freedom in BZTO. 

\section{Conclusions}
BZTO, which is an indirect bandgap semiconductor, undergoes a structural transition from high-symmetry $R\bar{3}m$ phase to low-symmetry $C2/m$ phase with the application of pressure at around 18~GPa. A thorough analysis of Raman spectroscopic data and DFT analysis show that the structural transition is lead by the softening of the low frequency N$_1$ (E$_{1g}$) phonon mode. All the Raman mode frequencies show a slope change around 7.5~GPa. The FWHM of the Raman modes decrease with pressure up to about 10~GPa and then increase due to anharmonicity corroborated by the over-damped soft mode. A sudden increase in trigonality along with the jump in bond angle variance of the TeO$_6$ octahedra introduce an increase in lattice-distortion. This confirms that the anomalies observed in the Raman modes are related to the structural instability developed in the parent structure as external pressure is applied. Absorption spectroscopic data and DFT calculations reveal that the sample is an indirect bandgap semiconductor with a bandgap of around 3.52~eV. The rate of increase of bandgap in $R\bar{3}m$ phase is 55 times more responsive to pressure compared to $C2/m$ phase. While a 44\% drop in the compressibility across the rhombohedral to monoclinic phase transition pressure indicates a strong correlation between the electronic and structural dynamics in BZTO. Continuous research on the structural and electronic relationship under pressure in the Ba$_2$M$^{2+}$Te$^{6+}$O$_6$ family remains essential as it paves the way for improving our understanding of these materials which in turn leads the development of materials with optimized properties for future technologies.
\section{Acknowledgments}
The authors acknowledge the financial support from the Department of Science and Technology, Government of India, to perform the experiment under the DST-DESY project in P02.2 extreme beam line condition at PETRA III, Germany. BM acknowledges the UGC, Government of India, for the
financial support to carry out the PhD work. S.A. would like to acknowledge the Council of Scientific and Industrial Research (CSIR), Government of India [Grant No. 09/1128(11453)/2021-EMR-I] for financial support. The authors acknowledge the High Performance Computing Cluster (HPCC) `Magus' at Shiv Nadar Institution of Eminence for providing computational resources that have contributed to the research results reported within this paper.

\newpage

\bibliographystyle{unsrt}
\bibliography{manuscript}

\newpage 

\section{Figures}		

%


\end{document}